# Unexpected Marangoni Condensation in Negative Binary Mixtures

*Abenezer Abere [a], Patricia B. Weisensee [a,b,*]*

[a]Department of Mechanical Engineering & Materials Science, Washington University in St. Louis, St. Louis, Missouri 63130, USA

[b]Institute of Materials Science and Engineering, Washington University in St. Louis, St. Louis, Missouri 63130, USA

*Corresponding author: 1 Brooking Drive, St. Louis, MO 63130, USA.

E-mail addresses: a.abenezer@wustl.edu (A. Abere), p.weisensee@wustl.edu (P.B. Weisensee) *.

## Abstract

Marangoni condensation—where surface tension gradients induce instabilities that lead to condensate film breakup into discrete droplets—has traditionally been considered limited to “positive" binary mixtures, where the less volatile component has higher surface tension. "Negative" mixtures were expected to exhibit stable filmwise condensation. Here, we demonstrate unexpected spontaneous Marangoni-driven pseudo-dropwise condensation in “negative” water–ethylene glycol and water–triethylene glycol mixtures. Strong thermo-diffusion in these dilute mixtures enables preferential glycol enrichment in colder condensate film regions during condensation, generating surface tension gradients that trigger film breakup, leading to >6× wettability-independent heat transfer enhancement compared to filmwise condensation at moderate subcooling. Our work challenges the conventional framework that restricts Marangoni condensation to "positive mixtures”—a superficial classification that oversimplifies the underlying interfacial mechanisms that can trigger robust Marangoni condensation, offering new pathways for enhancing phase change heat transfer in industrial applications without the need for expensive and degradation-prone surface coatings.

# 1. Introduction

Condensation, a ubiquitous phase change phenomenon, plays a central role in power generation, heating, ventilation, and air conditioning (HVAC) systems, thermal management, desalination, and atmospheric water harvesting technologies [1–5]. The efficiency of these processes is determined by the condensation mode. In filmwise condensation, a thick liquid layer covers the entire condensing surface, which acts as thermal resistance that hinders efficient heat transfer [6,7]. In contrast, dropwise condensation is characterized by discrete droplets that continually nucleate, grow, and shed the surface by gravity or vapor shear [7,8]. Dropwise condensation can enhance heat and mass transfer by more than an order of magnitude compared with its filmwise counterpart [7,8], but usually requires surface functionalization [1,9–13]. Although initially effective, the applied coatings face significant challenges in durability and reliability [14–18], making them unsuitable for long-term industrial applications [12,14,15,19].

Alternatively, condensation enhancement can be achieved by leveraging interfacial instabilities that arise from the fluid properties of multicomponent fluid systems rather than surface modifications. During the condensation of vapor mixtures, naturally occurring temperature and composition inhomogeneities can create surface tension gradients along the condensing liquid interface [20–23]. For most binary mixtures, surface tension is particularly sensitive to composition at low concentrations of the secondary component, typically $0.005 - 5\ wt\%$ [22–27]. Slight variations in interfacial composition can hence lead to strong surface tension gradients that can induce spontaneous Marangoni convection within the condensate film and initiate or enhance interfacial instability [20–23,28]. This instability can grow and break the condensate film into a set of discrete droplets, independent of the substrate's wettability [20–22]. This Marangoni-driven instability is a passive and self-sustaining mechanism that resembles a quasi-dropwise condensation mode without the need for substrate modifications. Despite the fundamentally different mechanism of droplet formation compared to traditional dropwise condensation, Marangoni condensation can sustain up to 10-fold enhancements in heat transfer performance compared to filmwise condensation; very similar to wettability-driven dropwise condensation [21,22,24,26,29,30].

Historically, the occurrence of Marangoni condensation of binary vapors has been explained by the relationship between the surface tension and volatility of the two components and the resulting sign of the surface tension gradient on the condensing film [22,23]. Generally, the change in surface tension with respect to the film thickness must be positive for any film instabilities to grow and ultimately to break up into individual droplets. It is thought that a mixture must be "positive" (for example, a water-ethanol mixture), meaning the surface tension of the high boiling point component (*e.g.*, water) must be higher than the surface tension of the component with the lower boiling point (*e.g.*, ethanol) [23]. Some azeotropic mixtures can also locally satisfy the instability criterion [26,31]. In contrast, a "negative" non-azeotropic mixture (*e.g.*, water-ethylene glycol), for which the surface tension of the high boiling component (ethylene glycol) is lower than that of the more volatile component (water), is thought to stabilize the film and impede any occurrence of pseudo-dropwise Marangoni condensation [21–23].

In this work we show that this historical framework is incomplete. Through a combination of experimentation and macroscopic interfacial thermodynamic modeling using water-glycol mixtures as exemplary "negative" systems, we propose that thermo-diffusion drives local compositional gradients that destabilize the condensate film, triggering spontaneous and robust droplet formation, motion, and shedding, irrespective of the substrate wettability. When a temperature gradient is imposed on dilute water-glycol mixtures, as is the case for condensation on a cooled substrate, thermo-diffusion drives the glycol molecules from warmer toward colder regions, enriching the thinner (and cooler) areas of the uneven condensate film with glycol [28,32–34]. This migration changes the interfacial composition and modifies the local surface tension, such that compositional gradients may oppose stabilizing thermal gradients, leading to the growth of film instabilities and eventual breakup into individual droplets [21–23]. Thus, we show that even systems nominally classified as "negative" mixture can exhibit spontaneous Marangoni condensation. These findings reveal a new pathway for achieving sustained pseudo-dropwise condensation without being limited to the use of surface coatings or the traditional "positive" mixtures by leveraging intrinsic molecular coupling effects in binary mixtures.

# 2. Materials and Methods

## 2.1 Mixture Preparation

Binary vapor mixtures of water–ethylene glycol and water–triethylene glycol were prepared at a 100:1 volume ratio, unless otherwise noted. All reported mixture compositions refer to the initial liquid composition used to prepare the mixture. Reagent-grade ethylene glycol (EG) and triethylene glycol (TEG) (both Reagentplus, ≥99%, Sigma-Aldrich) were used as received. For most experiments, the condensing substrate was a polished flat copper sample with a root mean square roughness of $R_q \approx 62.1$ nm, as characterized by atomic force microscopy (AFM; Bruker MultiMode) in ScanAsyst mode over a 10 μm by 10 μm scan area and analyzed using the NanoScope Analysis 3.00 software. Further details are provided in Sections S1 and S2 of the Supplementary Material (SM).

## 2.2 Condensation Visualization

Condensation visualization experiments were performed using an experimental setup housed in a fume hood in the presence of non-condensable gases (Fig. S1). Vapor generated in a cylindrical stainless-steel chamber equipped with immersion cartridge heaters was directed onto the copper substrate mounted vertically against a cold plate maintained at 25°C by a recirculating chiller (VWR Scientific 1175PD). Condensation dynamics were recorded using a DSLR camera (Canon T6i) equipped with a macro lens (Canon MP-E 65mm, 1–5x magnification). The progression of condensate film formation, instability growth, and droplet shedding was documented through time-lapse imaging. Full details of the experimental setup and visualization procedure, condensation observation on the different substrates, and corresponding wettability characterization are provided in Sections S1–S3 of the SM.

## 2.3 Contact angle measurement

To assess the influence of substrate wettability on the observed condensation behavior, advancing and receding contact angles ($\theta_a$ and $\theta_r$) and the contact angle hysteresis of the condensing binary mixture on smooth polished copper, polished aluminum, hydrophilic glass, hydrophobic PDMS, and superhydrophobic nanoparticle (Glaco) substrates were measured using a contact angle goniometer (Model 250-U4, ramé-hart) with a volume addition and withdraw method. The contact angles were measured by advancing or receding $\sim 5\mu l$ droplets on the surfaces using a syringe pump. Measurements at four different locations on each surface were averaged, with the associated standard deviations reported as uncertainties.

**2.4 Vapor–Liquid Equilibrium and Surface Tension Modeling**

Vapor–liquid equilibrium (VLE) phase diagrams for the water–EG and water–TEG systems were computed using COSMO-RS (AMS 2025.1) [35], which combines quantum-chemically derived molecular surface charge densities with a statistical-thermodynamic framework to predict activity coefficients and mixture phase equilibria. A comprehensive explanation of this model can be found in refs. [36,37]. The VLE was evaluated at 1 bar for both water-EG and water-TEG systems to match the conditions for the condensation visualization experiments, and additionally at 53 kPa = 0.53 bar for the water-EG mixture to match the operating pressure of the heat transfer experiment. Additional details can be found in Section S4 of the SM. The composition- and temperature-dependent mixture surface tension was calculated as proposed by Shardt *et al.* [38], evaluated along each systems' saturated liquid line at its corresponding operating pressure (SM Section S5). The sensitivities of surface tension to composition $(\partial\sigma/\partial x)$ and to temperature $(\partial\sigma/\partial T)$ $(\partial T/\partial x)$ along the liquid-vapor interface were evaluated numerically along the saturated liquid line of the respective VLE. Details of the thermodynamic modeling and sensitivity analysis are provided in Sections S4–S6 of the SM.

## 2.5 Heat Transfer Measurements

Condensation heat transfer experiments were conducted in a custom-built stainless-steel chamber (Fig. S7) equipped with a 1.5" sapphire viewport, a high-accuracy capacitance manometer (728A13TGA2FA Baratron, MKS) for saturation pressure monitoring, and a high-capacity recirculating chiller (Thermochill II, Thermo Scientific) delivering coolant to a vertically mounted copper condensing block thermally isolated from the vapor by a polytetrafluoroethylene (PTFE) housing. Chilled water impacted the back side of the copper block, which consisted of several equally spaced cylindrical fins to enhance the heat removal process and maintain the test sample at the desired temperature.

The condensation heat transfer coefficient was determined from a one-dimensional Fourier conduction analysis through the instrumented copper block using three embedded 10 kΩ NTC thermistors (803- USP10975, Mouser electronics), with the condensing surface temperature extrapolated from a fourth thermistor embedded in the test surface and the measured heat flux. The experiments were conducted over a subcooling range of $5.6 \leq \Delta T \leq 23$ K with condensation dynamics captured by an infrared camera (Telops M3k) through the sapphire viewport. The accessible subcooling range was constrained, in part, by the temperature control system. The high-capacity recirculating chiller has a maximum coolant setpoint of 30 °C, making very low subcooling temperatures experimentally inaccessible. At the highest measured subcooling temperatures, the experimental apparatus was limited by the vapor supply stability. The error for the condensation heat transfer coefficient was determined by propagating the uncertainty of the individual measurement contributions. The classical Nusselt filmwise condensation model, validated against a pure-water experiment, served as the baseline. Full details on the setup, procedure, calculations, and uncertainty analysis are provided in Sections S7–S10 of the SM.

# 3. Results and Discussion

## 3.1 Mechanistic Understanding of Marangoni Condensation

Initially, Marangoni condensation starts out similar to filmwise condensation, that is, with the formation of a thin condensate film covering the cooled substrate, within which small thickness variations naturally arise [22,23]. As heat is transferred from the vapor through the thin liquid film to the cooled substrate, these thickness variations lead to a temperature gradient along the interface, whereby thinner troughs are slightly colder and thicker ridges slightly warmer due to the thickness-dependent conduction resistances between the hot vapor and cold substrate, as shown in Figure 1(a). Heat diffuses across the film of thickness $\delta$ on a timescale $\delta^2/\alpha$ (thermal diffusivity $\alpha$), whereas species diffuse on a timescale $\delta^2/D$ (molecular diffusivity $D$). The ratio of thermal diffusion to species diffusion is described by the Lewis number, $Le = \alpha/D = 1.5 \times 10^{-7}\ m^2s^{-1}/(6 \times 10^{-1}\ \ m^2s^{-1}) \sim 100$ for water-EG [33], indicating that thermal conditions generally are established significantly faster than composition fields. The film composition therefore responds to an already established temperature distribution. However, where each component of the binary mixture subsequently preferentially condenses and accumulates within the perturbed film—at the colder troughs or warmer crests (see Figure 1(a))—depends on the mixture's thermophysical properties and vapor–liquid equilibrium.

Once the interfacial temperature gradient set by the varying film thickness is established, the local interfacial composition follows from the mixture's vapor–liquid phase diagram and the concept of interfacial partitioning: the colder troughs become enriched in the more volatile component, *i.e.*, the component with higher vapor pressure, and the warmer ridges in the less volatile component, as seen in Figure 1(a) [22,23]. Hence, a composition gradient develops along the liquid film. This compositional gradient, in turn, induces a surface tension gradient, which drives Marangoni flow of the interfacial liquid from lower surface tension region to higher surface tension. The location of the lower surface tension (troughs *vs*. ridges) therefore dictates the direction of this flow and, ultimately, whether the film becomes unstable. For water–ethanol (positive mixture), ethanol has the lower surface tension and higher volatility and preferentially accumulates at the colder troughs, locally reducing the surface tension there (see Figure 1(a)) [22,23]. The resulting Marangoni flow consequently drives liquid from the troughs toward the higher surface tension crests, which leads to film instability. This instability grows and eventually breaks the film and results in the formation of individual droplets.

The onset of instability can be expressed by the necessary, but not sufficient condition [23]:

$$\partial\sigma(T,x_1)/\partial\delta \equiv \left(\partial\sigma(T,x_1)/\partial T\right)_{sat}\left(\partial T/\partial\delta\right) > 0, \qquad \text{Eq. (1)}$$

where $\sigma(T,x_1)$ is the local surface tension of the liquid mixture, which depends on the local interfacial temperature $T$ and the interfacial liquid mole fraction of the minor component $x_1$, $\delta$ is the local film thickness, and the subscript *sat* represents local saturation conditions. For the cooled condensing surface, $\partial T/\partial\delta > 0$ due to conduction through the film. Therefore, $\partial\sigma(T,x_1)/\partial T$ must be greater than zero to satisfy the instability condition in Eq. (1) for Marangoni condensation to occur.

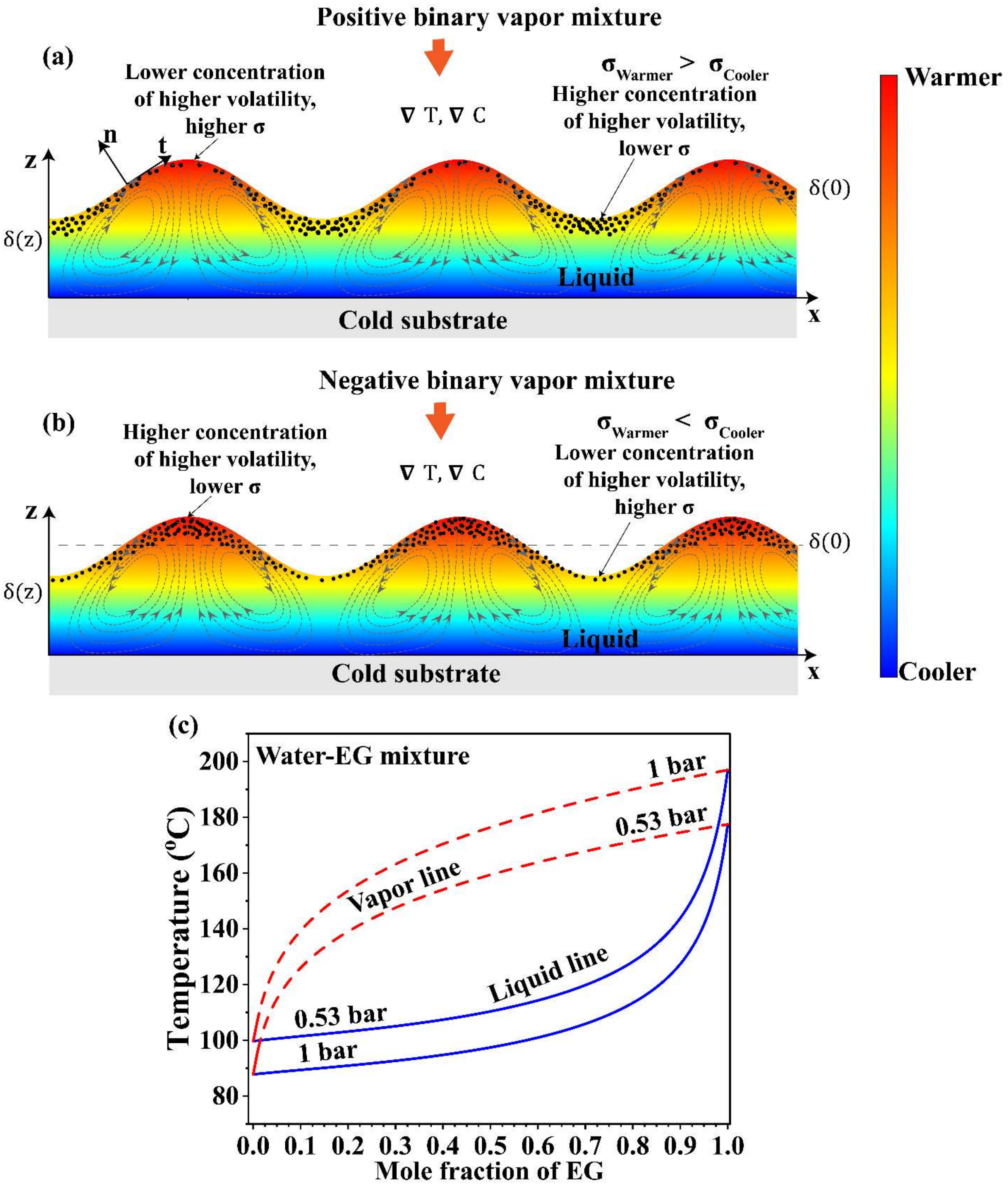

***Figure 1***. *(a) Sketch of a perturbed thin condensate film for a "positive" system , such as a water-ethanol mixture: The resultant temperature gradient causes the colder valley region to have a higher concentration of the more volatile component, in this case ethanol, than the warmer ridge, leading to a lower surface tension in the cooler region* ($\sigma_{valley} = \sigma_{cooler} < \sigma_{ridge} = \sigma_{warmer}$)*. The surface tension gradient leads to Marangoni flow from the valley to the ridge,*

*enhancing the fluid dynamic instability, until the film eventually breaks up to form droplets. (b) Sketch of the historical understanding of a perturbated thin film for a "negative" system, such as water-ethylene glycol. The temperature gradient should cause the warmer ridge to have a higher concentration of the less volatile ethylene glycol than the colder valley, leading to a lower surface tension on the ridge ($\sigma_{valley=cool} > \sigma_{ridge=warme}$). Marangoni convection should hence transport liquid from the ridge to the valley, stabilizing the film. Sketches are not drawn to scale. (c) Phase equilibrium diagram of the water-EG mixture. The solid blue lines indicate the liquid line and the red dashed lines indicate the vapor line.*

In "negative" systems like water–ethylene glycol mixtures, the less volatile component has the lower surface tension; in this case ethylene glycol, which – according to the VLE – now preferentially accumulates on the warmer ridges (see Figure 1(b-c)). Consequently, the water-rich colder regions are expected to have a higher surface tension than the warmer EG-rich regions. According to the conventional stability criterion Eq. (1), this should stabilize the condensate film, as the surface tension gradient opposes the direction needed for instability growth and Marangoni convection points from the ridge to the valley. Hence, no Marangoni condensation should occur.

However, contrary to the expectation, water–ethylene glycol and water–triethylene glycol mixtures do exactly that: they condense in a pseudo-dropwise manner, as shown in Figure 2 for condensation of water–EG and water–TEG mixtures at a 100:1 volume ratio on an uncoated copper substrate. Details on these visualization experiments are given in the SM, Sections S1-S3, and supplementary videos S1 and S2.

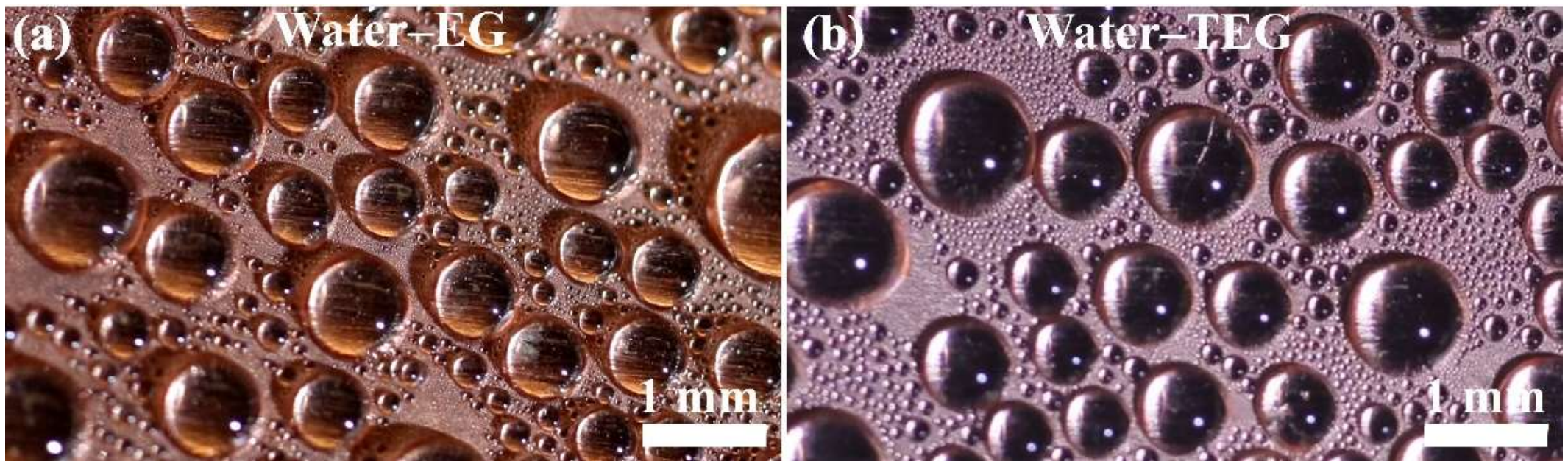

***Figure 2**. Representative images of fully developed Marangoni-driven pseudo-dropwise condensation modes of (a) a dilute water–EG mixture and (b) a dilute water–TEG mixture, respectively. Subcooling temperature $\approx 12K$ in an open laboratory environment.*

These observations raise a fundamental question: Why does such an obvious and pronounced Marangoni response emerge in a negative system that should, according to the traditional framework, favor stable filmwise condensation?

Unlike many aqueous alcohol systems, water–glycol mixtures exhibit a strong coupling between solute-water interactions and thermo-diffusion (*i.e.*, Soret effect) [32,34,39]. We propose that the interplay between compositional gradients and thermal diffusion within the perturbed condensing film leads to spontaneous condensate film instabilities in these "negative" binary systems, eventually leading to Marangoni condensation in a system that traditionally has been thought of inevitably exhibiting filmwise condensation on high-surface-energy metallic substrates. To support our hypothesis, we next examine how coupled variations in temperature and composition modify the surface tension and what physical mechanism initiates the surface tension gradients necessary to drive the observed instability growth.

### 3.2 Surface Tension Modeling Based on the Vapor Liquid Equilibrium

To determine the origin of the hydrodynamic instabilities in water–glycol systems, which still have to satisfy the instability criterion of Eq. (1), *i.e.*, $\partial\sigma/\partial T > 0$, but clearly do not follow the traditional "volatility *vs.* surface tension" framework, we first turn to mapping the interfacial thermodynamic states for a water-ethylene glycol mixture using interfacial temperature and composition information from the vapor-liquid equilibrium diagram (see Figure 1(c)). Next, we calculate the corresponding temperature- and composition-dependent surface tension of the water–EG mixture ($\sigma_{mix}$) with [38] :

$$\sigma_{mix} = \sigma_2(T) - \left(1 + \frac{b(1 - x_1)}{1 - a(1 - x_1)}\right) x_1\big(\sigma_2(T) - \sigma_1(T)\big), \qquad \text{Eq. (2)}$$

where $x_1$ denotes the liquid-phase mole fraction of the non-aqueous component ($0 \leq x_1 \leq 1$) and $\sigma_1(T)$ and $\sigma_2(T)$ are the temperature-dependent surface tensions of the non-aqueous component and pure water, respectively. The dimensionless parameters $a$ and $b$ are obtained by fitting existing surface tension data as a function of composition at a given reference temperature [40].

Water–EG and water–TEG mixtures exhibit a monotonic dependence of surface tension with both temperature and concentration (see SM, Fig. S6). This trend reflects the thermodynamic trajectory of the mixture, since temperature and concentration co-vary along the phase equilibrium path rather than acting as independent variables. To quantify and compare the relative contributions of temperature and composition on surface tension, we evaluate the sensitivity of surface tension to composition $(\partial\sigma/\partial x)$ and to temperature $(\partial\sigma/\partial T)\,(\partial T/\partial x)$ along the vapor liquid equilibrium, as shown in Figure 3 (see SM, Section S6 for details on modeling). Both contributions have the same units, allowing their magnitudes and signs to be compared directly. The compositional contribution is the product of the surface-tension sensitivity to composition and the change in liquid composition with temperature at saturation equilibrium. We can identify two regimes: the dilute and the concentrated regime. Generally, surface tension is more sensitive to changes in composition than to temperature variations [38]. This is more pronounced in the dilute regime, where slight compositional changes sharply change surface tension compared to higher concentration regions ("concentrated regime") where effects are less pronounced, indicating that the solutal effect is dominant for our condensation experiments. We define $x_1 \approx 0.05$ as an approximate boundary for the dilute region due to the highest sensitivity of surface tension to compositional changes (and, as will be discussed later in Section 3.3, the peak of the effective surface tension). Both composition and temperature sensitivities plateau and remain nearly constant as the composition increases in the concentrated regime, with changes in concentration always having a stronger influence on surface tension than temperature variations.

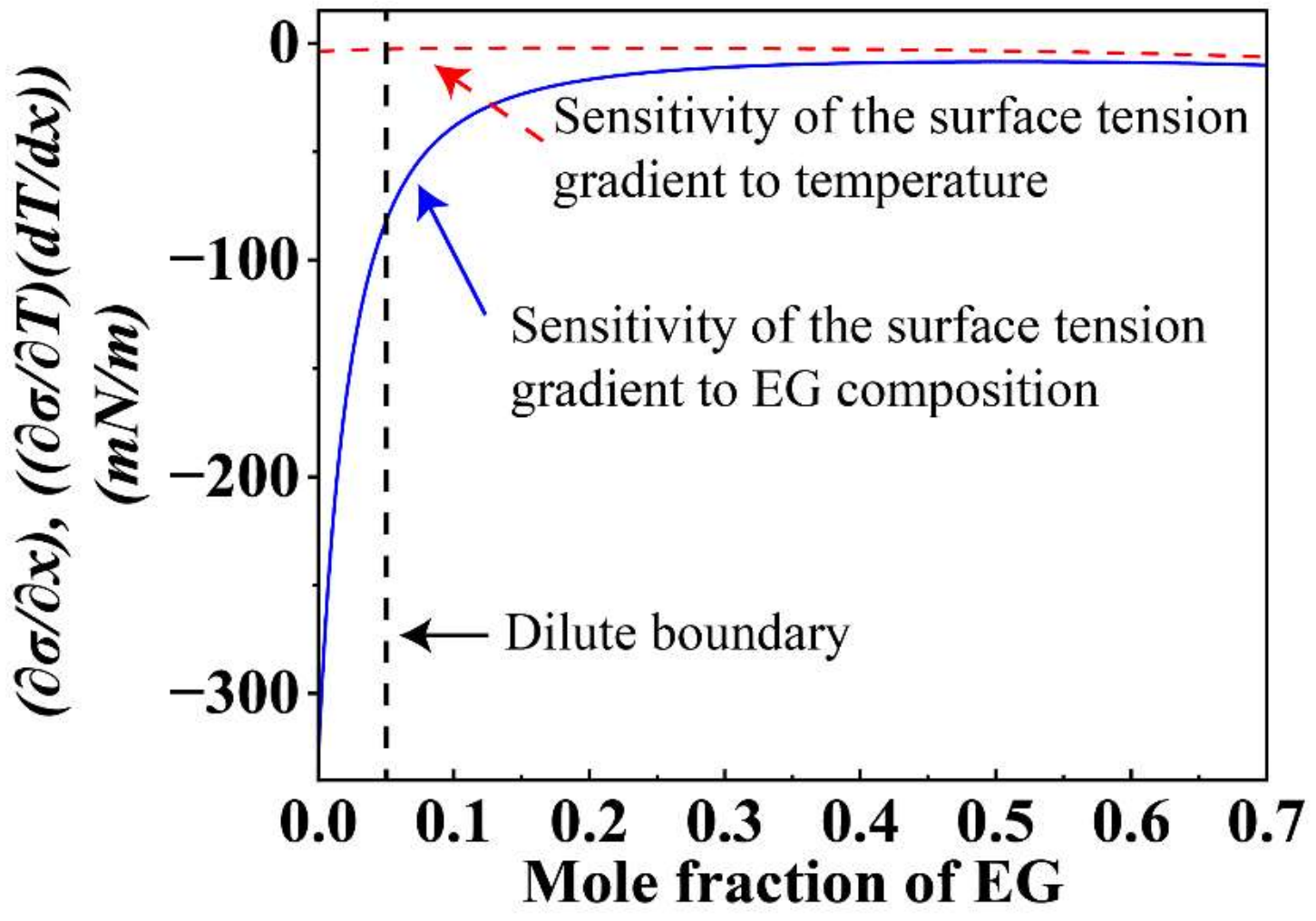

***Figure 3.*** *Sensitivity of the surface tension gradient to composition and temperature examined through their respective partial derivates ((*$\partial\sigma/\partial x$*) and* $(\partial\sigma/\partial T)\,\partial T/\partial x$*), respectively) for a water-EG mixture. The solid blue line represents the relative magnitude change of the surface tension response to composition, while the dashed red line shows its response to temperature. The vertical broken line differentiates between the dilute and the concentrated regimes. The units of the partial derivatives are mN/m per unit mole fraction.*

The pronounced solutal dominance observed in Figure 3 aligns with the underlying molecular mechanisms governing surface tension in the dilute regime. In this region, composition variations disrupt the hydrogen-bond network far more effectively than temperature changes, making solutal effects the principal drivers of interfacial tension gradients and, therefore, interfacial flow. This strong sensitivity of surface tension to composition in the dilute regime can be attributed to distinct molecular interactions, specifically the competition between water-solute and water-water hydrogen bonding [32,41]. These interactions can alter or break the hydrogen bonding between water and solute molecules (EG) [32,39,41–44], leading to significant changes in interfacial thermodynamic properties [41,45,46]. There are two primary mechanisms by which hydrogen bonds between water and solute molecules can be disrupted in aqueous mixtures [32,46]. First, the addition of a solute with strong affinity for water, such as EG [43], restructures the water hydrogen bond network by forming extensive hydrogen-bonded complexes with water molecules [32,46].

Second, increasing temperature disrupts hydrogen bonding interactions [47]. At higher concentrations, the mixture adopts a new hydrogen-bonded structure, characteristic of glycol-rich phases, reducing the sensitivity of surface tension to composition and temperature.

The local variation in hydrogen bonding—driven by the composition and temperature inhomogeneity—directly control the interfacial surface tension. Any spatial composition non-uniformity induced by the temperature gradients in the condensing film will thus generate surface tenson gradients that drive interfacial film instability [22,23]. Hence, this raises the question of which physical mechanism initiates the surface tension gradients necessary to drive the onset of the observed instability during condensation.

**3.3 Molecular Transport in Water-Glycol Mixtures Leads to Instability Growth**

When a water-EG mixture is subjected to a temperature gradient, the molecules do not merely conduct heat, they also migrate due to thermo-diffusion, creating concentration gradients [32,34,48,49]. This phenomenon is quantified by the Soret coefficient, $S_T$, which characterizes the ratio of thermal diffusion to molecular diffusion in the system [48,49]. The resulting concentration gradient can be expressed as:

$$\nabla c = -S_T\, x_1(1 - x_1)\nabla T, \qquad \text{Eq. (3)}$$

where $x_1$ is the mole fraction of solute component (EG) and $\nabla T$ is the existing temperature gradient. The direction of species migration is determined by the sign of the Soret coefficient: positive values indicate solute migration toward colder regions, whereas a negative value indicates solute migration toward hotter regions.

As established in the previous section, EG initially preferentially condenses on the warmer film ridges, in accordance with the interfacial partitioning deduced from the VLE (see Figure 1(c) and Figure 4(a)). Within the film, however, the temperature gradient also drives EG migration due to thermo-diffusion. Water–glycol mixtures show complex and anomalous thermodynamic behavior compared with simpler aqueous alcohols [39,41,50]. In the dilute regime of the water-EG mixture, the Soret coefficient is positive and strongly composition dependent, with remarkably high values at low EG concentrations [33,34]. This behavior is closely tied to how these mixtures respond to temperature gradients, leading to rich and often pronounced thermo-diffusion effects. A positive Soret coefficient means that EG migrates from warmer to colder regions. Thermo-diffusion therefore transports EG from the warmer ridges toward the colder troughs, where it then accumulates (see Figure 4(b)) [32–34]. Under an imposed temperature field and in the diffusion-controlled limit, the resulting concentration redistribution develops on a characteristic timescale of , $\delta^2/D$, across the film, which is of the order of milliseconds for the micrometer-thick films typically observed in Marangoni condensation [51,52].

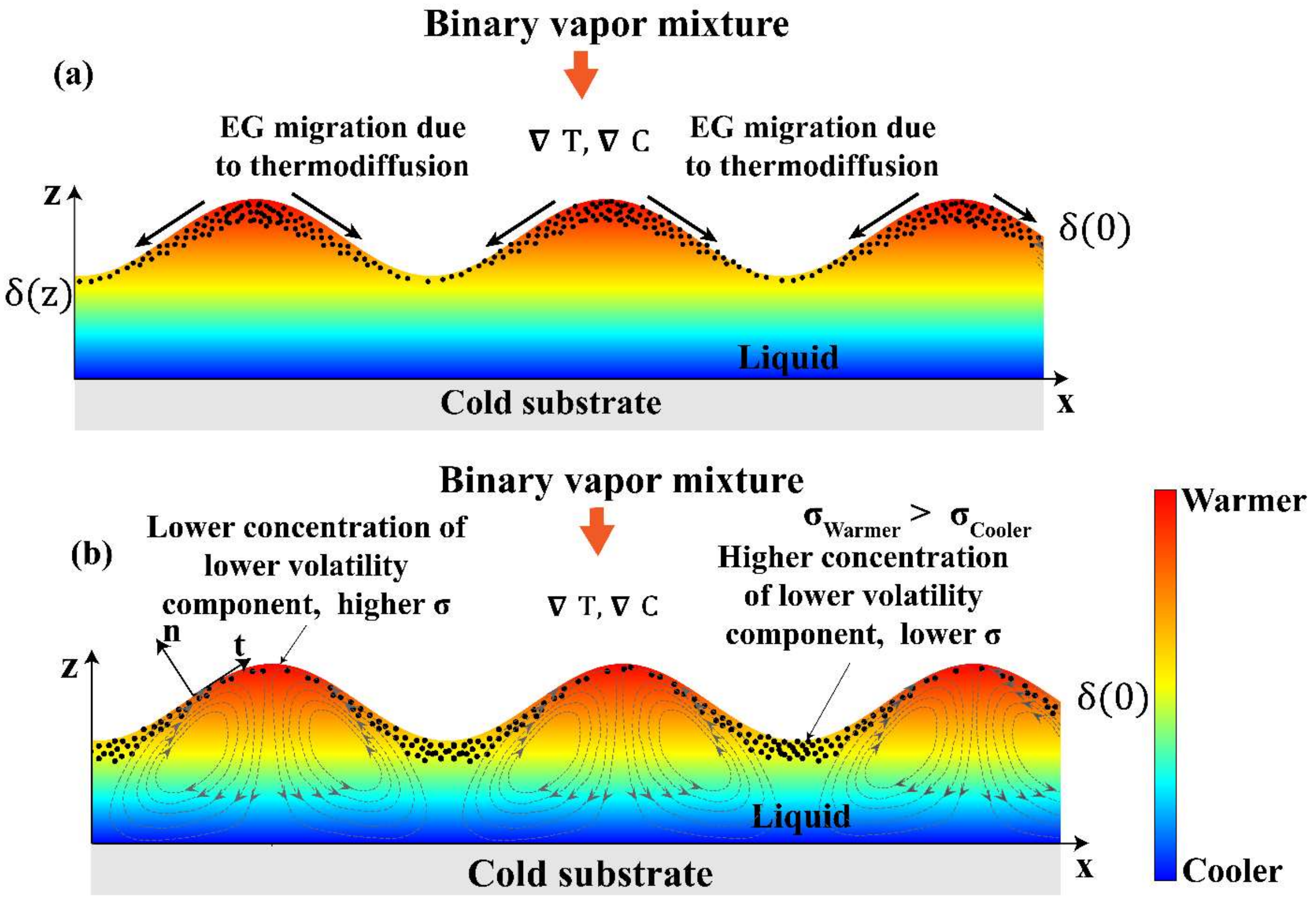

***Figure 4*** *(a) Sketch of a thin perturbed film showing the early stages of water-EG condensation, where EG initially accumulates on the peak region due to its lower volatility compared to water and then migrates towards the cooler region due to thermo-diffusion. (b) Sketch of a perturbed film showing the accumulation of EG in the cooler region due to thermo-diffusion, resulting in the positive surface tension gradient that triggers the film instability to grow. Sketches are not drawn to scale.*

The resultant solute redistribution thus competes with the thermal profile to reverse the sign of the surface tension gradient: the colder trough, which for single-component liquids would have the higher surface tension than the warmer ridge, becomes more enriched with EG, which effectively lowers its surface tension due to the high sensitivity of surface tension on concentration, as seen in Figure 3: small composition gradients can induce large surface tension gradients.

To demonstrate this sign reversal, we define an effective surface-tension gradient, $\sigma_{eff} = d\sigma/d\delta$, as:

$$\sigma_{eff} = \sigma_{\mathrm{T}}(T, x_1) - \sigma_{\mathrm{C}}(T, (x_1 + \nabla c_S)), \qquad \text{Eq. (4)}$$

where the corresponding values of $\sigma_{\mathrm{T}}$, the temperature-dependent surface tension gradient at the reference composition and $\sigma_{\mathrm{C}}$, the Soret-modified surface tension gradient are computed using Eq. (2) along the phase diagram shown Figure 1(c). The Soret-induced compositional redistribution ($\nabla c_S$) is evaluated using Eq. (3), which requires an approximation of the temperature gradient ($\nabla T$) along the interface. Here, we assume the temperature gradient between the crest and trough to be 2K, consistent with a previously reported value during Marangoni condensation [25]. Data for the composition-dependence of the Soret coefficient, $S_T(x_1)$, is taken from reference [33] and incorporated into the calculation of $\nabla c_S$ across the composition range. To facilitate the estimation of the coupled thermal and solutal effects, local vapor–liquid equilibrium is assumed at the interface (to use the VLE), while acknowledging that heat and species transport within the condensate are non-equlibrium processes. Thermo-diffusion can redistribute EG within the film and thereby influence the interfacial composition through the coupled species balance. Here, the estimated Soret redistribution is used to assess whether compositional changes could reverse the calculated surface-tension difference. This provides a screening criterion for the direction of Marangoni flow rather than a complete prediction of film instability.

The sign of $\sigma_{\mathrm{eff}}$ indicates the direction of the effective surface tension gradient and consequently the direction of flow. For $\sigma_{\mathrm{eff}} > 0$, surface flow is from the trough to the ridge, amplifying the surface perturbations, whereas $\sigma_{\mathrm{eff}} < 0$ indicates flow from the crest to the valley, stabilizing the film. Evaluating this coefficient across a range of bulk liquid compositions (*i.e.*, $x_1$) provides a sign map of the thermo-solutal response, as shown in Figure 5(a). It can be seen that the sign of $\sigma_{\mathrm{eff}}$ for the dilute regime remains positive. In this range, thermo-diffusive redistribution of the EG lowers the surface tension in the colder troughs and opposes the stabilizing thermal gradient. We therefore conclude that the system within these compositional ranges is susceptible to Marangoni-driven convection that can trigger and sustain film instability, leading to film break-up and droplet formation, with the strongest activity occurring around $x_1 \approx 0.05$, which we used to define the dilute regime cut-off in section 3.2. As the composition of EG increases, however, the sign of $\sigma_{\text{eff}}$ reverses near $x_1 \approx 0.2$, and beyond this point the compositional redistribution reinforces, rather than opposes, the thermal gradient. Hence, molecular re-distribution no longer overcomes thermal stabilization, and the system is predicted to be stable and demonstrate a filmwise condensation mode.

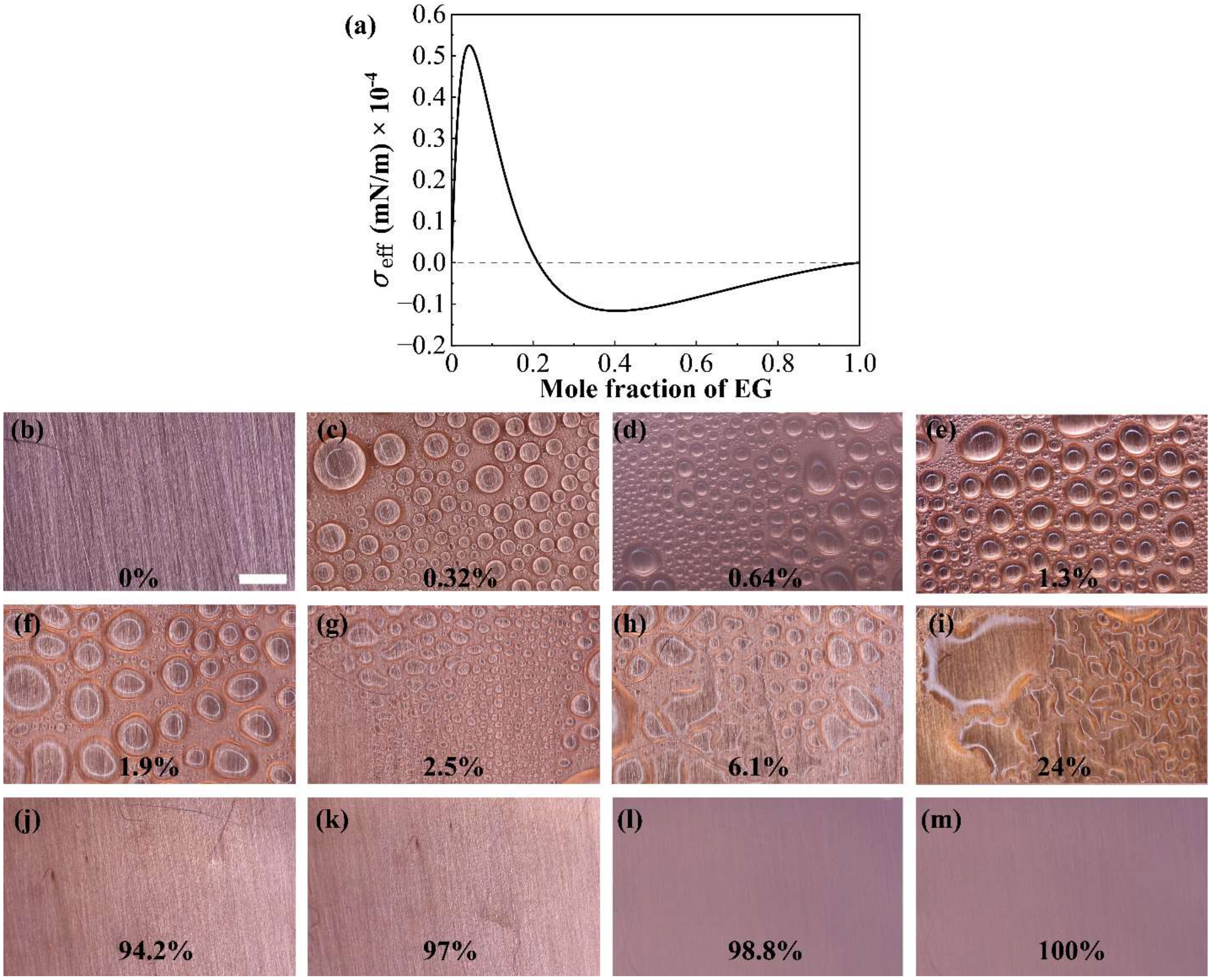

***Figure 5.*** *Prediction map of the effective surface tension gradient and observed condensation mode across the water-EG composition range. (a) Effective surface tension gradient ($\sigma_{eff}$) as a function of EG mole fraction. (b-m) Representative condensation images on polished copper ($\theta_a/\theta_r = 88.1 \pm 3.1/34.4 \pm 5.5$) for (b) pure water, (c-i) dilute EG mixtures with (EG mol%) $\approx 0.32, 0.64, 1.3, 1.9, 2.5, 6.1, 24$ (j-l) EG-rich mixtures (EG mol%) ≈ 94.2, 97, 98.8, and (m) pure EG. The scale bar is 1mm and applies to all images.*

To validate this model, we examined the predicted composition dependence by performing condensation experiments across a concentration series under the same experimental conditions (see Figure 5(b-m)). Pure water and pure EG, for which binary compositional redistribution is absent, both formed continuous films, as expected on highly wetting copper surfaces. Dilute mixtures containing 0.32–6.1 mol% EG (Figure 5(c-h)), associated with predicted $\sigma_{\rm eff} > 0$, formed discrete droplets with a broad size distribution. In contrast, the mixture around $x_1 \approx 0.24$ (Figure 5(i)), close to the predicted sign change around $x_1 \approx 0.20$ displayed a transitional film morphology and the EG-rich mixtures containing 94.2–98.8 mol% EG (*i.e.*, 1.2-5.8 mol% water) remained filmwise as well, highlighting that the phenomenon and critical mixture concentrations are species-dependent and not applicable to any dilute mixture. We thus demonstrate that thermo-diffusion along with the coupled thermal-compositional influence on surface tension can have dramatic impacts on the evolution of the condensation process of carefully selected binary mixtures.

Finally, to assess the influence of substrate wettability, the condensation behavior was additionally visualized on polished aluminum, hydrophilic glass, hydrophobic PDMS, and superhydrophobic nanoparticle substrates. Together with polished copper, these surfaces span a broad wettability range, with measured apparent contact angles ranging from $33°$ to $152°$ and contact-angle hysteresis values of approximately 1°–54°. Pseudo-dropwise condensation was observed on all investigated substrates, demonstrating its occurrence across this wettability range. Representative condensation images and surface wettability measurements are provided in Fig. S(2-3) of the SM.

### 3.4 Enhanced Condensation Heat Transfer

Having established that composition gradients in the water-glycol mixture can induce condensate film instability and trigger Marangoni driven dropwise condensation, we next assessed whether this mechanism leads to measurable enhancements in thermal performance by examining the heat transfer coefficient as a function of wall subcooling. The goal was not to provide comprehensive heat transfer data, but rather to demonstrate that pseudo-dropwise condensation of "negative" mixtures has similarly heat transfer enhancing properties as that of "positive" mixtures.

To demonstrate the improvement of the condensation heat transfer performance, a water-EG mixture was selected as a representative system and tested in a custom-built condensation chamber (see SM sections S7 through S10 for details). Prior to performing the experiment, the vapor was generated by rigorously boiling the mixture to remove non-condensable gases. A high-speed infrared (IR) camera was used to image the condensation process in front view through a sapphire viewport. Time-lapse IR images obtained during the Marangoni dropwise condensation of a water-EG mixture on the clean smooth copper surface show discrete droplet formation, coalescence and gravity-driven shedding (See SM Fig. S11, video S3 and video S4).

Figure 6(a) shows the measured condensation heat flux as a function of surface subcooling ($\Delta T$) for an exemplary, non-optimized ≈0.55 $wt\%$ (liquid phase; $x_1 \approx 0.16\, mol\%$) water–EG mixture undergoing Marangoni dropwise condensation on a smooth copper surface. The heat flux decreased from approximately $630\,\mathrm{kW/m^2}$ at a subcooling of $\Delta T = 5.6$ K to a minimum of approximately $490\,\mathrm{kW/m^2}$ near a subcooling of 16 K. This decrease may be attributed to weakening surface-tension variations along the condensate interface with increasing subcooling, which reduces the Marangoni driving force that maintains the droplet-like condensate [29,53]. At $\Delta T \approx 17$ K, the heat flux increased slightly to ≈590 $\mathrm{kW/m^2}$ and then remained nearly constant, while the heat transfer coefficient in Figure 6(b) shows a corresponding plateau before resuming its decrease. This change coincides with an accidental increase in the vapor inflow during the experiment. Higher vapor velocities thin the diffusion layer near the interface, reducing the vapor-side resistance and raising both the heat flux and the heat transfer coefficient [24,29].

Figure 6(b) shows the measured condensation heat transfer coefficient decreases monotonically with increasing condenser subcooling over the accessible subcooling range $5.6 \leq \Delta T \leq 23$ K. At the lowest measured subcooling, $\Delta T = 5.6$ K, the condensation heat transfer coefficient reaches $114.8 \pm 9.7\ \mathrm{kW/m^2K}$, corresponding to an enhancement of roughly $6.7 \times$ compared to filmwise condensation of pure steam ($\approx 17\ \mathrm{kW/m^2K}$). This enhancement arises from the reduced thermal resistance during Marangoni dropwise condensation, where discrete droplets form, coalesce, and shed *via* gravity, sweeping adjacent droplets to expose fresh surface areas which enable the repeated formation of new droplets. As the subcooling increases, the heat transfer enhancement decreases, yet remains greater than unity even at the highest subcooling tested ($\Delta T \approx 23$ K), demonstrating a broad operational window in which Marangoni-driven dropwise condensation consistently outperforms filmwise condensation, even when the driving temperature difference becomes large.

At even higher subcooling temperatures, which are inaccessible to our experimental setup, the heat transfer enhancement associated with Marangoni condensation may diminish as the condensate evolves toward larger, more irregular droplets and rivulets [26,29,52]. With increasing subcooling, the dominant thermal resistance shifts toward the condensate layer, while the interfacial surface tension gradients that sustain the Marangoni-driven condensate motion become less effective in redistributing the growing amount of condensate, resulting in a decline in heat transfer performance [26,29,52]. It is well known that a further increase in subcooling eventually leads to a transition toward a smooth condensate film [26,29,52].

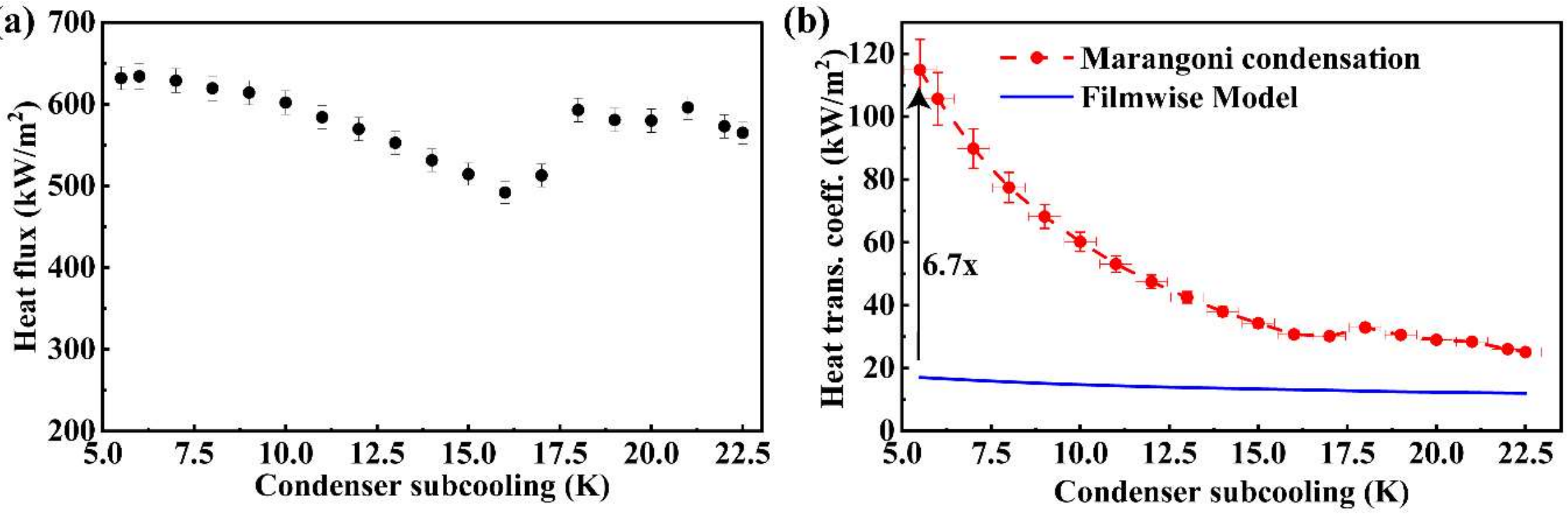

***Figure 6.*** *Enhanced condensation heat transfer performance. Heat flux (a) and condensation heat transfer coefficient (b) of a dilute water-EG mixture (≈0.16 mol%) as a function of condenser subcooling temperature at a vapor pressure of* $52.86 \pm 1.62\ kPa$. *The theoretical prediction (solid blue line in (b)) represents the Nusselt model for filmwise condensation of pure quiescent water vapor on a bare copper condenser and serves as a baseline case for comparison with the measured condensation heat transfer performance (for model derivation, setup validation and parameters, see SM section S10). The error for the measured condensation heat transfer coefficient was determined by propagating the uncertainty of components associated with measurement (described in detail in the SM section S9). The kink in the data around* $\Delta T = 17K$ *is due to a slight increase in vapor inflow.*

## 4. Conclusion

This work challenges the conventional framework that limits Marangoni condensation to "*positive mixtures*"—a superficial classification that oversimplifies the interfacial liquid-vapor transport physics [20–22].

We show that in "negative" water-EG and water-TEG mixtures, the subtle interfacial coupling of compositional and temperature gradients *via* thermo-diffusion drives the preferential migration of glycol species towards colder regions [32,34], reducing the mixture's surface tension in the film's troughs. The resulting surface-tension gradient generates the Marangoni stresses necessary to destabilize the condensate film, triggering pseudo-dropwise condensation in systems traditionally expected to form stable films [20–22]. The condensation mode observed for these water-glycol mixtures resembles that of traditional dropwise condensation on wettability-modified surfaces [7,8,11,12] and classical Marangoni condensation of "positive" mixtures, such as water-ethanol [20,22,23,29,30]. Similarly, we demonstrate a more than 6x condensation heat transfer performance enhancement compared to filmwise condensation, without the need for any surface modifications, offering new pathways for enhancing condensation performance in industrial applications.

Discovering the influence of thermo-diffusion on pseudo-dropwise condensation shifts the currently prevailing description of Marangoni condensation from a purely volatility-driven phenomenon to one governed by coupled interfacial transport physics. More broadly, these findings reveal that Marangoni condensation is applicable to a wider range of systems beyond traditional "*positive mixture*" criteria. Building upon the foundational insights presented in this work, we recommend that future research quantify the limits of surface tension differences, thermo-diffusion coefficients, and volatility through a mathematical analysis of the phase change and molecular diffusion dynamics. We encourage further exploration of additional molecular mechanisms that might contribute to Marangoni condensation, particularly the thermo-physical anomalies of water, such as shifts in hydrogen bonding strength and surface activity in dilute aqueous mixtures [54,55].

**CRediT authorship contribution statement**

**Abenezer Abere:** Conceptualization, Data curation, Methodology, Validation, Formal analysis, Investigation, Software, Writing – original draft, Visualization. **Patricia Weisensee:** Conceptualization, Formal analysis, Resources, Writing – review & editing, Supervision, Project administration, Funding acquisition.

## Acknowledgment

This material is based upon work supported by the National Science Foundation under Award No. 2147483.

# Unexpected Marangoni Condensation in Negative Binary Mixtures

Abenezer Abere [1], Patricia B. Weisensee [1,2*]

*[1] Department of Mechanical Engineering & Materials Science, Washington University in St. Louis, St. Louis, Missouri 63130, USA*

*[2] Institute of Materials Science and Engineering, Washington University in St. Louis, St. Louis, Missouri 63130, USA*

Supplementary Material

*Author to whom correspondence should be addressed.

Patricia B. Weisensee: p.weisensee@wustl.edu.

## Contents

## S1. Experimental setup for Marangoni condensation visualization

Figure S1 shows a schematic of the experimental setup that was used for obtaining imaging data (heat transfer and thermal data used a different setup, as described in section S7), consisting of a vapor generator, a cold stage, and an imaging system. The entire setup was housed in a fume hood and the experiments were conducted in the presence of non-condensable gases. This setup was employed for testing whether the binary vapor mixtures would undergo Marangoni condensation. The binary mixtures were prepared according to the ratio specified in the manuscript and physical properties of the corresponding substances are summarized in Table S1. The volume ratio was converted to mass fraction using $w = (\rho_{EG}V_{EG})/(\rho_{EG}V_{EG} + \rho_{water}V_{water})$.

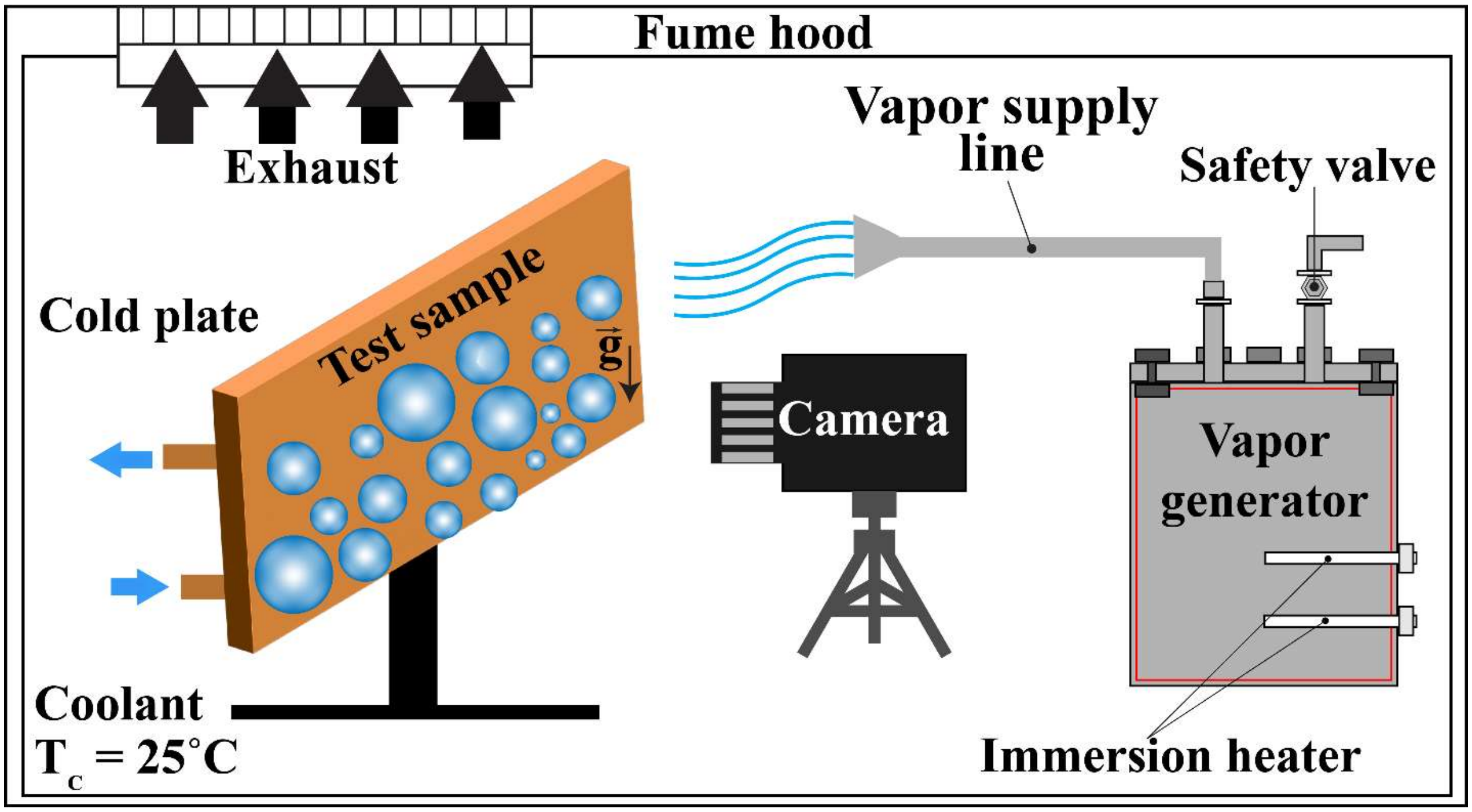

**Figure S1.** *Schematic of the experimental setup for Marangoni condensation demonstration and visualization. Schematics are not drawn to scale.*

*Table S1. Physical properties of substances used in the experiment*

| **Substance** | **Supplier** | **Density (kg/m$^3$)** | **Molar mass (g/mol)** | **Surface tension (mN/m) at 20°C** |
|---|---|---|---|---|
| DI water | | 997 | 18.015 | 72.8 |
| Ethylene glycol (Reagentplus, ≥ 99%) | Sigma-Aldrich | 1,113 | 62.07 | 48.4 |
| Triethylene glycol (Reagentplus, ≥ 99%) | Sigma-Aldrich | 1,124 | 150.17 | 45.2 |

A cylindrical stainless-steel chamber equipped with two immersion cartridge heaters (400W, OEM Heaters) served as a vapor generator. Two separate KF adapter ports were installed on the top of the chamber. The first port was used to attach a bellow angle valve (Kurt J. Lesker) to the vapor generator, which served as a safety vent for over-pressure protection. The second port was connected to a flexible tube that served as the flow line for the incoming vapor mixture, which was channeled towards a condensation surface. A flat copper sample was mounted vertically and secured against the front side a tubed cold plate to serve as condensation surface. A thin layer of thermal paste was applied between the copper sample and the cold plate to ensure good thermal contact. The tubed cold plate was connected to a recirculating chiller (1175PD Recirculating Chiller, VWR Scientific) to supply cooling water at a constant temperature of 25 ℃. A DSLR camera (Canon T6i), equipped with a macro lens (Canon MP-E 65mm 1-5x Macro), was used to record the condensation dynamics. While heat transfer rates are expected to differ greatly depending on the atmosphere (open lab *vs*. environmental chamber without non-condensable gases) due to the distribution of thermal resistances (gas boundary layer *vs*. liquid film), the presence of non-condensable gases does not necessarily suppress Marangoni-driven droplet formation [1].

## S2. Surface characteristics and wettability

Figure S2 and Table S2 show and list the sessile, advancing ($\theta_a$), and receding ($\theta_r$) contact angles of a 0.32 mol% water-EG mixture on the five tested substrates, respectively.

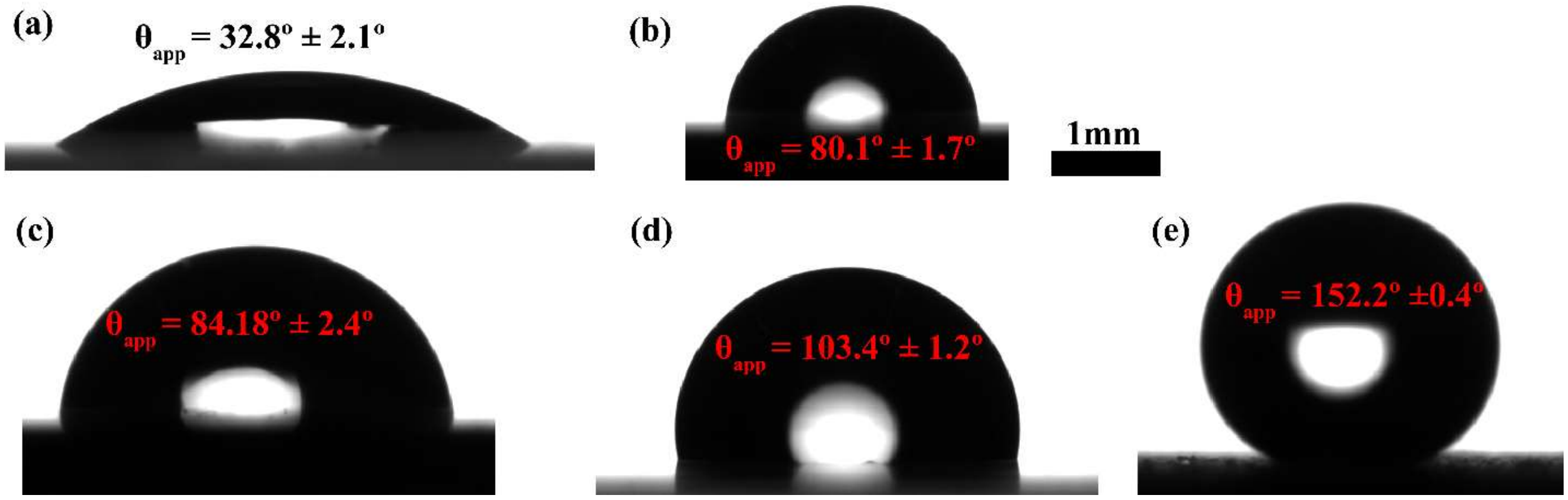

**Figure S2.** *Representative images showing the sessile contact angles of a 0.32 mol% water-EG mixture on the investigated substrates: (a) hydrophilic glass, (b) polished aluminum, (c) polished copper, (d) hydrophobic PDMS, and (e) superhydrophobic nanoparticle surface.*

*Table S2. Advancing and receding contact angles and contact-angle hysteresis of the water–EG working mixture (0.32 mol% EG) on the five substrates. Values are mean ± standard deviation of four measurements.*

| Substrate | $\theta_a$ (°) | $\theta_r$ (°) | $\Delta\theta$ (°) |
|---|---|---|---|
| Hydrophilic glass | 43.1 ± 2.7 | 13.9 ± 2.7 | 29.2 ± 5.4 |
| Polished aluminum | 83.6 ± 2.1 | 29.5 ± 6.5 | 54.1 ± 8.6 |
| Polished copper | 88.1 ± 3.1 | 34.4 ± 5.5 | 53.8 ± 8.6 |
| Hydrophobic PDMS | 112.0 ± 1.5 | 100.2 ± 1.3 | 11.8 ± 2.7 |
| Superhydrophobic | 151.6 ± 0.5 | 150.3 ± 0.9 | 1.3 ± 1.4 |

Figure S3 shows exemplary images captured during the condensation of the water-EG mixture on various substrates having different surface wettability.

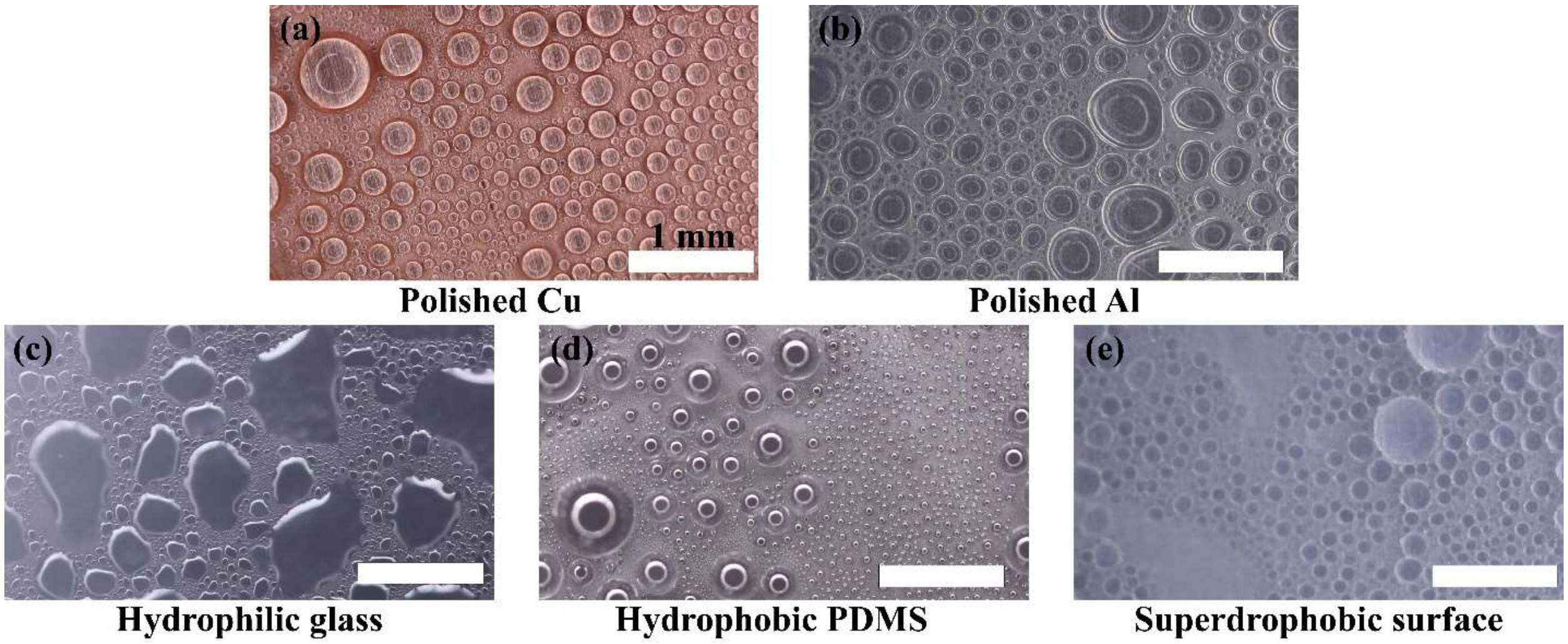

*Figure S3. Representative images showing steady-state condensation of a water-EG mixture on (a) polished copper, (b) polished aluminum, (c) hydrophilic glass, (d) hydrophobic PDMS and (e) a superhydrophobic nanoparticle surface. Scale bars are all 1 mm.*

Droplet-forming condensation was observed on both aluminum and copper despite their large contact-angle hysteresis, with predominantly fully developed droplets on both surfaces (Fig. S3(a,b)). On hydrophilic glass, a quasi-pseudo-dropwise condensation was observed, with droplets exhibiting highly irregular morphologies (Fig. S3(c)). In contrast, hydrophobic PDMS supported small, circular condensate droplets (Fig. S3(d)). The superhydrophobic surface exhibited a high advancing contact angle and low hysteresis for deposited droplets; however, condensate droplets remained pinned throughout the observation period (Fig. S3(e)), while remaining discrete without observable surface flooding during the observation period.

Overall, persistent pseudo-dropwise condensation was observed across all investigated substrates and wettability ranges, even though droplet morphology and pinning behavior varied among substrates.

Additionally, after condensation, the advancing and receding contact angles on the copper surface were $\theta_a = 87.6.0° \pm 2.0°$ and $\theta_r = 40° \pm 6.3°$, yielding a contact angle hysteresis $\Delta\theta \approx 47.6°$; comparable to the 53.8° measured before condensation. Although the hysteresis decreased modestly, it remained substantial, indicating that the surface retained appreciable contact-line pinning rather than becoming a low-hysteresis, slippery surface. These results suggest no significant change in wetting behavior following condensation.

## S3. Atomic force microscopy (AFM)

The roughness of the condensing copper substrate was characterized by AFM (Bruker MultiMode) using silicon nitride probes in the ScanAsyst mode. AFM images, as shown in Figure S4, were acquired by scanning areas of 10 μm by 10 μm on the copper surfaces in air under ambient laboratory conditions at a scan rate of 1 Hz. Two images were analyzed with NanoScope Analysis 3.00 software to obtain the root mean square roughness ($R_q$).

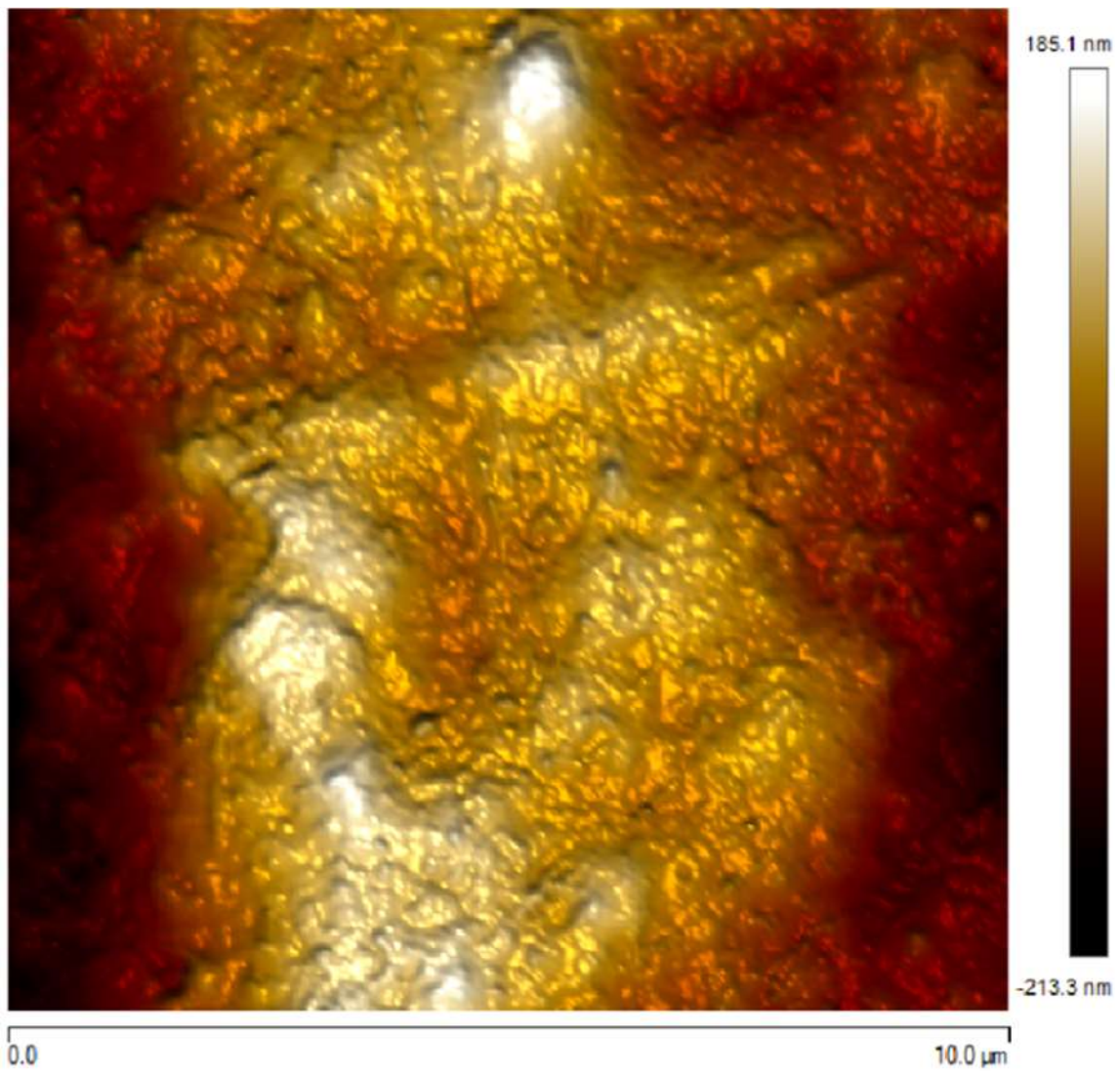

**Figure S4**. *AFM topography scan of the copper surface used for the experiment.*

## S4. Phase diagrams of binary mixture systems

The vapor liquid equilibrium (VLE) phase diagrams of the binary systems of water-ethylene glycol (W-EG) and water-triethylene glycol (W-TEG) are shown in Figure S5(c) of the main manuscript and Figure S5, respectively. The red dashed line indicates the vapor line whereas the blue solid line indicates the liquid line. The mixture thermodynamic properties were predicted using the COSMO-RS (Conductor-like Screening Model for Real Solvents) model implementation in the SCM Amsterdam Modeling Suite (AMS2025.1) [2]. COSMO-RS combines quantum-chemically derived molecular surface polarization charge densities with a statistical-thermodynamic treatment of pairwise surface-segment interactions to compute chemical potential and activity coefficients in liquids. These quantities are then used to predict mixture phase equilibria and related thermodynamic properties of binary mixtures. A comprehensive explanation of this model can be found in refs. [3,4].

The phase diagram shown in Figure S5 corresponds to the operating pressure of the condensation visualization experiments (Figure 2(b) of the main manuscript), which were conducted at 1 bar for both the water-EG and water-TEG mixtures. The heat transfer measurements reported in Fig. 6 of the main manuscript were conducted at the condenser operating pressure of 52.86 kPa (~0.53 bar) for the water-EG mixture. To ensure consistency, Figure S5(c) of the main manuscript therefore includes the water–EG phase diagram at 0.53 bar alongside the 1 bar case for comparison.

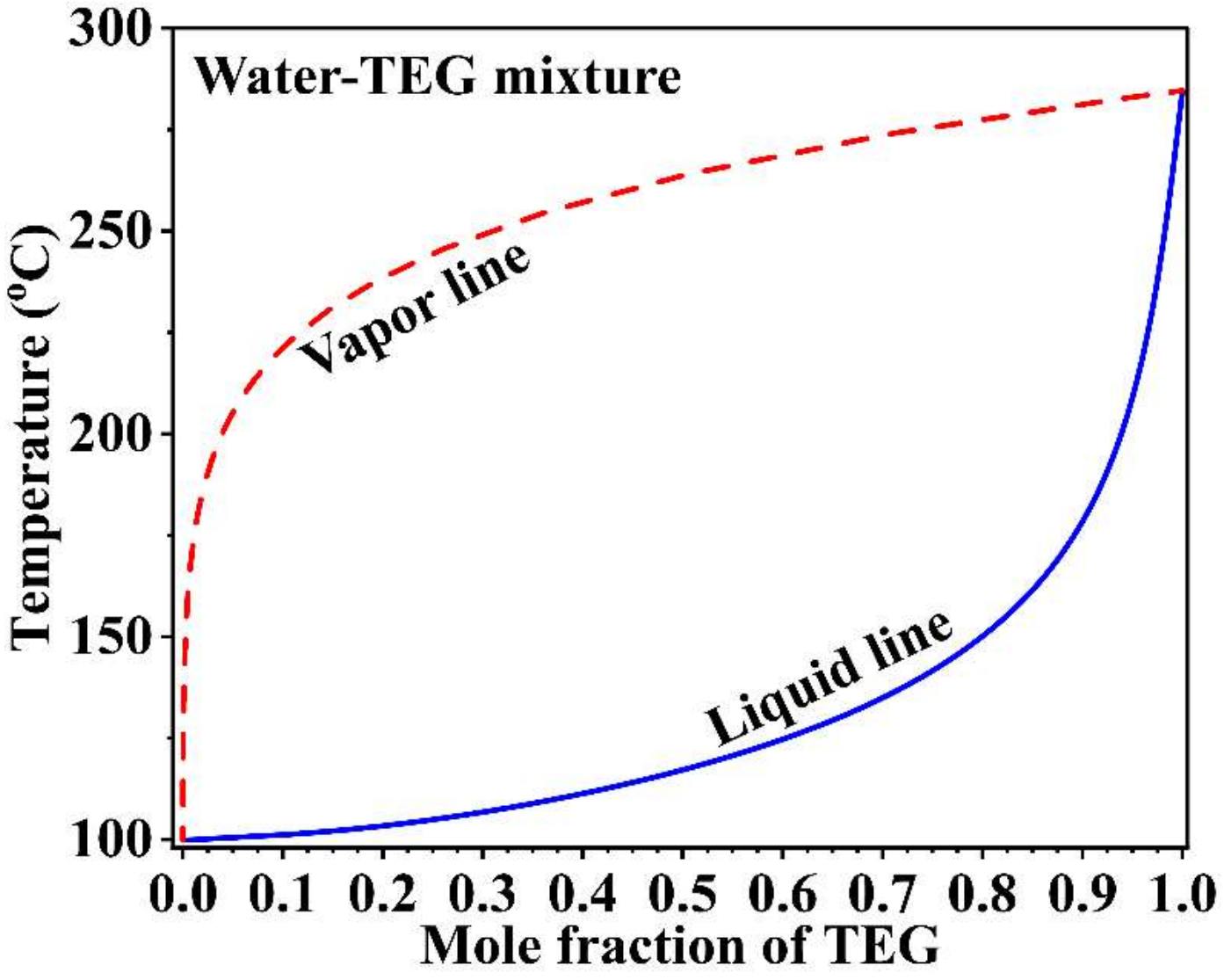

**Figure S5.** *Phase equilibrium diagram of the water-triethylene glycol mixture at 1 bar. The solid blue line indicates the liquid line and the red dashed line indicates the vapor line.*

COSMO-RS is a fully predictive (*a priori*) method, requiring no system-specific fitted parameters. Benchmarking against the standard phase equilibrium databases has shown that COSMO-RS performs well despite having an order of magnitude fewer adjustable parameters than the other models [5,6]. COSMO-RS has specifically been found to perform comparatively well for aqueous mixtures, the class of system relevant to this work [6].

Consistent with this, our comparison for water-EG system shows a good agreement with the experimental VLE data of ref. [7]. Restricting the comparison to the mole fraction of EG less than $x_1 = 0.20$, directly relevant to our experimental operating composition, we obtain a mean absolute deviation between our Cosmo-RS prediction and the data of ref. [7] of only 1.3K at 1 bar (*vs*. the closest experiment pressure, 747mmHg = 99.6 kPa).

We did not find available experimental data at our heat transfer operating pressure of 52.86 kPa; the closest reported experimental pressure is 430 mmHg (57.3 kPa) and we were therefore not able to make a direct comparison at our operating pressure. We also note that the dataset of ref. [7] reports only a limited number of compositions across the full range; the comparison should therefore be interpreted as indicating general consistency between our COSMO-RS predictions and experiment in this compositional range, rather than a finely resolved characterization at our exact operating composition and pressure.

## S5. Surface tension calculation of the binary mixtures

Once the composition and temperature data along the VLE is obtained from Figure S5(c) in the main manuscript for W-EG and Figure S5 for W-TEG, respectively, we use a modified form of the Connors–Wright model to calculate the surface tension of the binary mixtures, as proposed by Shardt *et al.* [8] and expressed as a function of liquid-phase composition and temperature:

$$\sigma_{mix} = \sigma_2(T) - \left(1 + \frac{b(1 - x_1)}{1 - a(1 - x_1)}\right) x_1\big(\sigma_2(T) - \sigma_1(T)\big), \tag{S1}$$

where $x_1$ denotes the liquid-phase mole fraction of the non-aqueous component ($0 \leq x_1 \leq 1$), $T$ is the absolute temperature from the VLE phase diagram, $\sigma_2(T)$ and $\sigma_1(T)$ are the surface tensions of pure water and the non-aqueous compound in (mN/m), respectively, and the dimensionless parameters $a = 0.95$ and $b = 0.79$ are obtained by fitting surface tension data [8,9] as a function of composition at a reference temperature chosen to have sufficient experimental points to capture the composition trend. The surface tension of pure water is calculated as [9]:

$$\sigma_2(T) = 235.8\left[\frac{T_c - T}{T_c}\right]^{1.256}\left[1 - 0.625\left(\frac{T_c - T}{T_c}\right)\right], \tag{S2}$$

where $T_c = 647.15K$ is the critical temperature and $T$ is the absolute temperature from the VLE phase diagram.

The surface tension of ethylene glycol is calculated as [10]:

$$\sigma_1(T) = 50.206 - 0.089T, \tag{S3}$$

where $T$ is in °C.

The surface tension of triethylene glycol is given by [11]:

$$\sigma_1(T) = 47.33 - 0.088T, \tag{S4}$$

where $T$ is in °C.

**Error! Reference source not found.** shows the calculated variation of surface tension as a function of saturation temperature along the liquid line of the vapor liquid phase equilibria. Note that composition also varies along the temperature axis, in accordance with the equilibrium conditions shown in the respective VLEs.

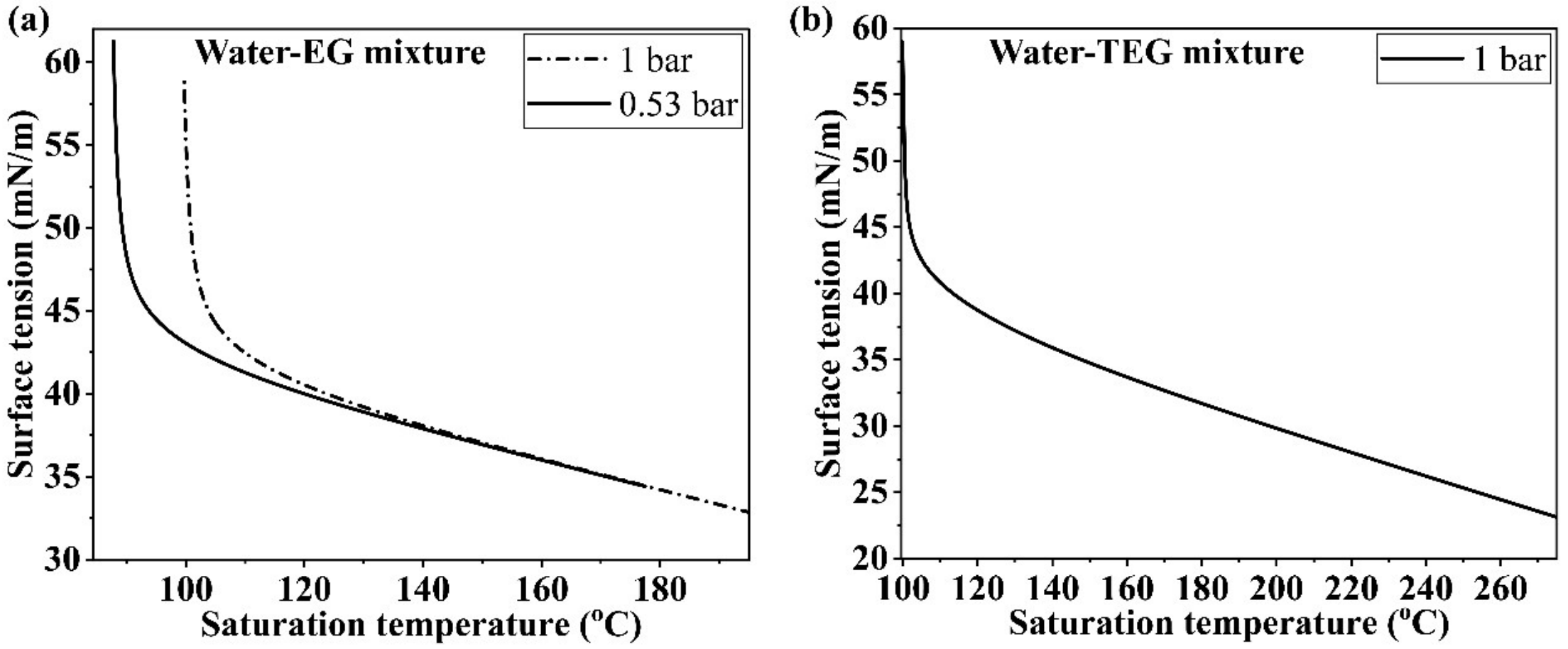

***Figure S6.*** *Surface tension of the mixtures as a function of equilibrium temperature (or liquid phase concentration) at the liquid-vapor interface. (a) Water-EG mixture at 1 bar (dash-dot line, visualization experiment condition) and ≈52.86 kPa, (solid line, heat transfer experiment condition). (b) Water-TEG mixture at 1 bar. The saturation temperature increase on the horizontal axis corresponds to the molar concentration increase of the solute according to the corresponding VLE phase diagram.*

## S6. Local sensitivity of surface tension

The surface tension of binary mixtures $\sigma_{mix}$, as defined in Eq. (S1), is a function of $T$ and $x_1$ at constant pressure, *i.e.*, $\sigma_{mix} = \sigma(T, x_1)$ [12]. To characterize the contributions of temperature and composition to its variation, the partial derivatives $(\partial\sigma_{\text{mix}}/\partial x_1)_T$ and $(\partial\sigma_{\text{mix}}/\partial T)_x$ were obtained by differentiating Eq. (S1). Both sensitivities were subsequently evaluated at the equilibrium temperature–composition pairs along the saturated-liquid line of the fixed-pressure VLE curve described in Section S4.

Because these partial derivatives have different dimensions, their numerical magnitudes were not compared directly. Instead, we can quantify the sensitivity of surface tension to changes in composition and temperature on a common dimensional basis using a chain-rule decomposition, expressed as follows [12]:

$$\left.\frac{d\sigma_{mix}}{dx_1}\right|_{\mathrm{VLE},P} = \underbrace{\left(\frac{\partial\sigma_{\mathrm{mix}}}{\partial x}\right)_T}_{C_x} + \underbrace{\left(\frac{\partial\sigma_{mix}}{\partial T}\right)_x \left.\frac{dT_{sat}}{dx_1}\right|_P}_{C_T}, \tag{S5}$$

where $T_{sat}$ is the saturation temperature, $C_x$ is the direct composition contribution, and $C_T$ is the temperature-mediated contribution along the equilibrium curve. The local slope, $(dT_{sat}/dx_1)_P$, was determined by numerical differentiation of the VLE data at each composition [12]. Both contributions therefore have units of $\mathrm{mN/m}$ per unit mole fraction, allowing their relative magnitudes to be compared using:

$$R = \frac{|\,C_x\,|}{|\,C_T\,|}\,. \tag{S6}$$

Values of $R > 1$indicate a larger composition contribution, whereas $R < 1$ indicates a larger temperature-mediated contribution to the surface-tension variation along the VLE curve. The data plotted in Figure 3 of the manuscript is calculated using eq. (S5) and a direct comparison of $C_x$ and $C_T$ is depicted.

## S7. Experimental setup for heat transfer measurements

We used a custom-made condensation chamber for heat transfer measurements with *in-situ* visualization capability, which consists of a vapor generator, a condensation chamber, a vacuum pump, a chilled water bath, and a data acquisition system, as shown in Figure S7.

A custom-built cylindrical stainless steel vapor generator (Ancorp), wrapped with two rope heaters (0.83kW and 0.62kW, Omega), was controlled by two independent Variac power controllers (3PN1010 Variac, Staco Energy Products Co.) and served as the vapor generator (see Figure S7). The vapor generator along with the rope heater were insulated externally to minimize heat loss to the environment. A stainless-steel bellow hose (Kurt J. Lesker), serving as the vapor supply line, was fed into the top of the vapor generator *via* a KF adapter port. The vapor supply line was wrapped with a rope heater (1.4kW, Omega), controlled by a Variac power controller (3PN1010 Variac, Staco Energy Products Co.), to prevent condensation at the inside wall and was thermally insulated to limit the heat loss to the environment.

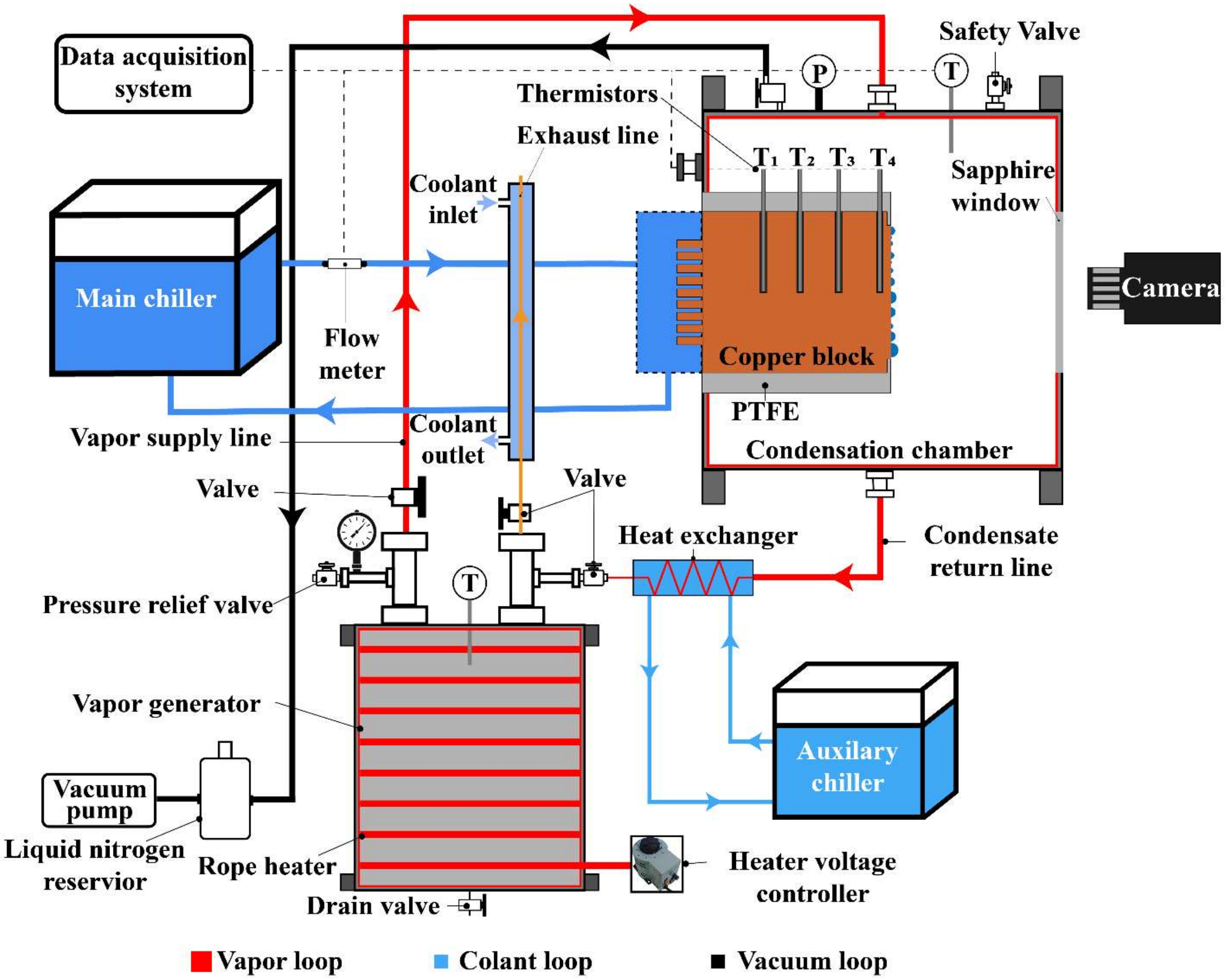

**Figure S7**. *Schematic of the experimental setup for heat transfer measurements. Schematics are not drawn to scale.*

A bellow valve (Kurt J. Lesker) was attached to the vapor generator *via* a KF port along the vapor supply line and was used to control the vapor inflow rate to the condensation chamber. A resistance temperature detector (RTD) probe (Evosensors) was fed into the vapor generator *via* a Swagelok fitting that monitored the temperature of the vapor generator. A T-fitting was installed on the vapor generator to integrate a pressure gauge (Bourdon needle dial, IdealVac) and a pressure-relief valve (Pfeiffer) *via* a KF port. The pressure inside the vapor generator was monitored by the pressure gauge, while the pressure relief valve served to protect the system from overpressure by venting excess vapor to the ambient environment.

A secondary stainless steel tube (exhaust line), integrated with an inline counterflow heat exchanger, was connected to the vapor generator. A vapor discharge valve (Ball valve, Idealvac) was installed along the line *via* a KF adapter port. The exhaust line served two purposes. First, it served as a port that was used to fill the liquid mixtures into the vapor generator through the vapor discharge valve. Second, during the boiling process, the vapor discharge valve was opened to allow the dissolved non-condensable gases to escape the chamber. The inline heat exchanger was connected to a secondary chiller (1175PD Recirculating Chiller, VWR Scientific) and used for condensing the vapor back into the vapor generator so that only the non-condensable gases were removed from the system during boiling. The outlet of the exhaust line was connected to a flexible tube, so that the non-condensable gases and a residue of the vapor escaping the condensing process were discharged into the fume hood. A drain valve was connected to the bottom side of the vapor generator *via* a Swagelok compression fitting to a stainless-steel tube that served as the drain port for the liquid mixture inside the boiler. Another stainless tube connected the bottom port of the condensation chamber (Swagelok) to the vapor generator and served as the condensate return line through gravitational suction. An auxiliary heat exchanger (Brazed Plate Heat Exchanger, Ferroday) was installed along the condensate return line to condense the remaining vapor back to a liquid and enhance the necessary suction.

The custom-designed condensation chamber used for this work (Ancorp) consisted of a cylindrical stainless-steel vessel horizontally seated on a stainless-steel stand (Ancorp) with a sapphire viewing window and apertures for various components. The vapor supply line was fed into the top of the condensation chamber *via* a KF adapter port to supply incoming vapor from the vapor generator. Flexible resistive heaters (Chromalox), powered by a DC power supply (Eventek), and insulation were wrapped around the exterior of the condensation chamber to prevent condensation at the inside wall. The vertical condensation copper block was thermally insulated from the vapor by a polytetrafluoroethylene (PTFE, McMaster-Carr) housing and was mounted against the rear side of the interior of the condensation chamber to serve as a condensation substrate. Two insulated water flow lines (Swagelok tee) were fed into the back side aperture *via* a CF flange (Kurt J. Leskter) to supply cooling water to the chamber from the main chiller (Thermochill II recirculating chiller, Thermoscientific). A flowmeter (Picomag, Endress+Hauser USA) with an accuracy of $\pm$0.8% was installed along the water inflow line.

A 4-way KF fitting (Kurt J. Lesker) was attached to one of the side ports of the condensation chamber. A high-accuracy pressure transducer (728A13TGA2FA Baratron Capacitance Manometer, MKS) was attached to top port of the 4-way KF fitting to monitor the gas and vapor pressure within the chamber and to determine the saturation pressure. The bottom port of the 4-way KF fitting was attached to a safety valve (Idealvac), which was used to release the chamber to ambient conditions if needed. An RTD probe (Evosensors) was fed through a Swagelok compression fitting on the opposite side of the condensation chambers and monitored the vapor temperature within the chamber.

One flange connected to a stainless-steel bellow hose (Kurt J. Lesker) that served as a vacuum line, which was connected to the vacuum pump (Adixen 2005 dual stage rotary vane, Idealvac) that was used to evacuate the condensation chamber prior to introducing the vapor. A ball valve (Idealvac) was integrated along the vacuum line to isolate the vacuum pump from the chamber and maintain the vacuum once the pump-down process was complete and the pump was turned off. A liquid nitrogen cold trap was also incorporated along the vacuum line to remove any moisture from the gas to protect the pump and to achieve a higher quality vacuum condition.

A 1.25” diameter polished flat copper condensing sample was vertically mounted and pressed against the copper cooling block substrate by the PTFE insulating block to ensure good thermal contact between the copper block and condensing sample. Thermal paste was applied between the interfaces to further improve heat transfer. The chilled water flowed through the water flow lines and impacted the back side of the copper block, which consisted of several equally spaced cylindrical fins to enhance the heat removal process and maintain the test sample at the desired temperature.

Bundles of thermistors were used to measure the temperature inside the condensation block, which was surrounded by the PTFE insulator. Three equally spaced 10 kΩ NTC thermistors (803-USP10975, Mouser electronics) were inserted into the condensing copper block to monitor the temperature. The corresponding temperature readings were used to calculate the 1D condensation heat flux through the copper block. A fourth thermistor (803- USP10975, Mouser electronics) was inserted into the condensing sample to extrapolate the surface temperature. All thermistors were wired to the electrical feedthrough CF flange (EFT0083032, Kurt J.Lesker) and wrapped with PTFE tape and then with shrink tube to shield them from the condensing vapor and minimize noise during measurements. The thermistors were calibrated by immersing them into a circulating bath (AD07R-40, Polyscience) filled with 50/50 water/ethylene glycol, which was held at a constant temperature to $\pm 0.01$°C. The thermistors in the water bath were allowed to reach steady state before collecting the calibration data. The readings from the thermistor (x-axis) and the actual/ true water bath temperature (y-axis) of Figure S8 were used as a calibration curve.

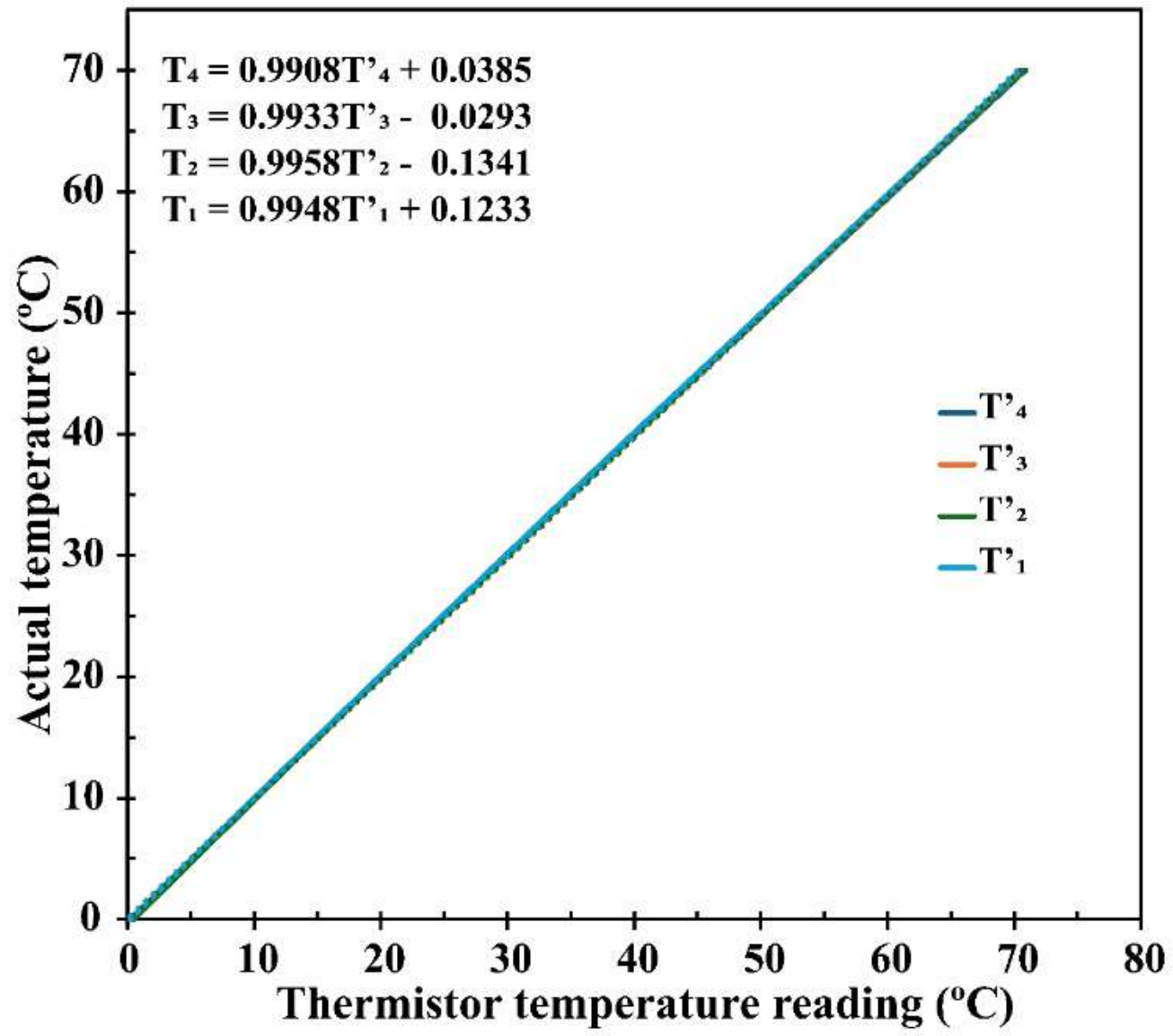

**Figure S8**. *Thermistor calibration curve. The bundle of thermistors was calibrated prior to the experiment by immersing them in a constant-temperature water/EG bath.*

The thermistor bundles and pressure transducer were electrically connected to the analog input source channel of a data acquisition system (Keysight technologies 34972A), which was interfaced to a computer for data recording. In order to visually record data, a high-speed camera (Photron Mini AX200, Japan) [13] was placed in line with the 1.5” sapphire viewing window on the chamber. In addition, a DSLR camera (Canon T6i) with a macro lens (Canon MP-E 65mm 1-5x SMacro), which was interchangeable with the high-speed camera, could be used to obtain color images. Moreover, since the sapphire window is infrared transparent, a thermal camera (Telops M3k) could replace the optical cameras to obtain spatial temperature information of the condensation process.

## S8. Experimental procedure

For each experimental trial, a set of strict procedures was followed to ensure consistency throughout the experiments. The first step was to ensure the environmental chamber was thoroughly clean and there were no traces of contamination from prior use. To do so, the chamber setup was filled with coca cola [14] for at least 8 hours and then drained from the chamber *via* a drain valve. Distilled water was then filled in the chamber and drained several times (8-10) to rinse the chamber and remove impurities. The next step of the process was to turn on the voltage regulators and power supply to heat up the environmental chamber wall to vaporize any residue of the distilled water used for cleaning and then remove it by turning on the vacuum pump. The hoses from the chamber were disassembled and the internal corrugated were dried using Kimwipes (Kimberly-Clark Professional). The hoses were left to air-dry for at least 24 hours before reassembly to ensure complete moisture removal.

Once the chamber was cleaned and re-assembled, prior to conducting condensation experiments, the voltage regulators and power supplies were turned on to heat up the environmental chamber to prevent condensation on the chamber walls. Next, the vacuum pump down procedure was initiated. First, the liquid nitrogen moisture trap was filled to about three-quarter capacity. The vapor supply valves connecting the vapor generator to the vapor supply line and the condensate return valve connecting the chamber to condensate return line were both fully closed. The ball valve connecting the chamber to the vacuum pump was opened. The vacuum pump was then turned on, initiating the pump-down process. The pressure inside the chamber was monitored during the pump-down process. This process took nearly four hours to achieve the best vacuum level our system can reach. Once the condensation chamber was pumped down to ≈280 Pa, the lowest pressure that could be reached in our system, the chamber was isolated from the vacuum pump and the pump was turned off. At one point, the chamber pressure was monitored overnight to characterize the leakage rate (Figure S9). The leak rate was 22.6 Pa/hr during the first 1.5 hours, then increased to approximately 40.5 Pa/hr for the remainder of the test duration.

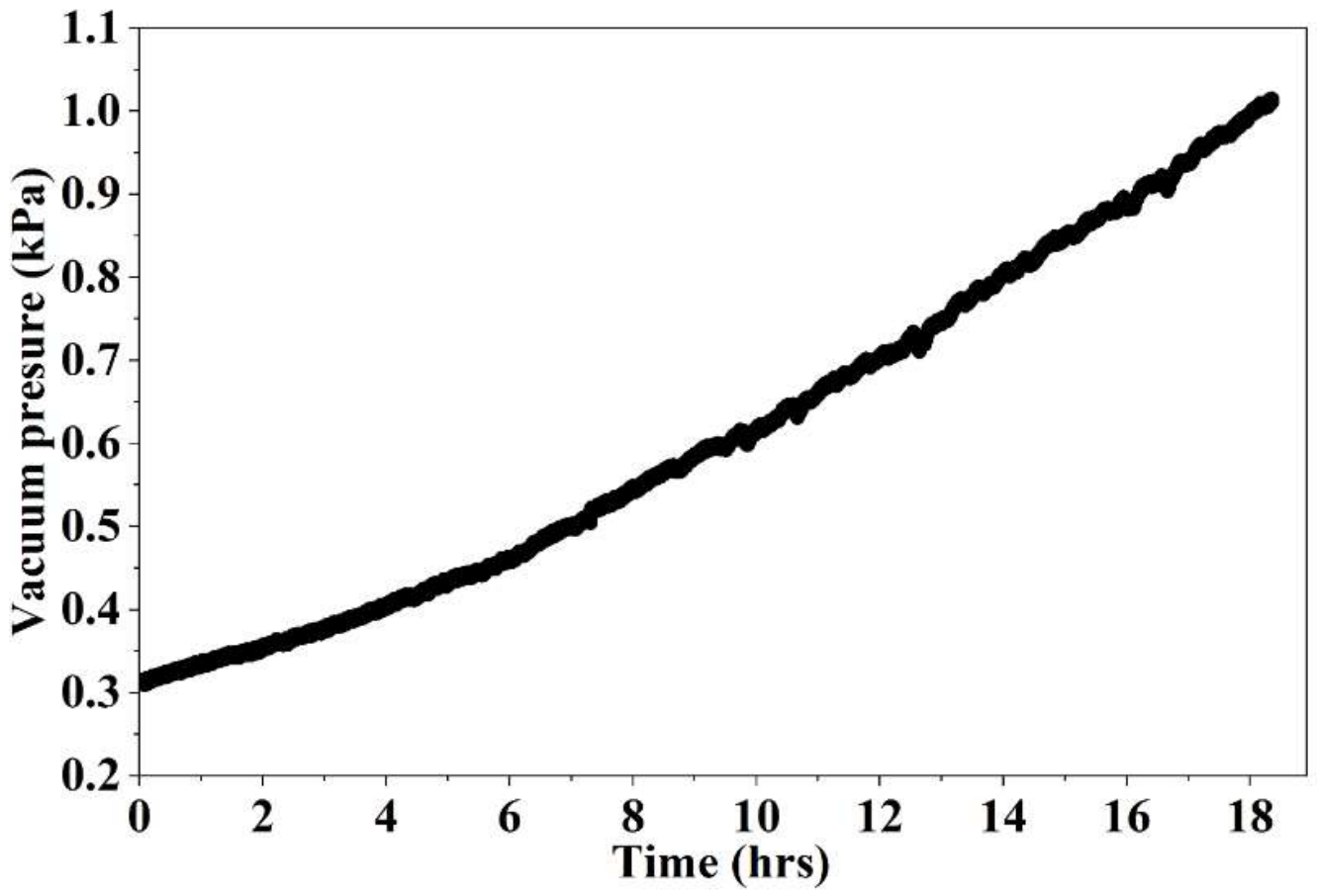

**Figure S9**. *Overnight leak rate characterization of the condensation chamber.*

During pump down, the setup procedure for the three water flow loops was performed simultaneously, as described below. A high-capacity chiller (Thermochill II recirculating chiller, Thermoscientific) was turned on in order supply the cooling water to the condensation copper block. The chiller output temperature was set between $-10$ °C and $30$°C to vary the condensing surface temperature, enabling the collection of heat transfer data over a range of subcooling conditions. The flowrate was monitored with the flowmeter integrated in the cooling water line. A secondary chiller (VWR Scientific), integrated with the parallel heat exchanger in the vapor discharge line that served for condensing back the vapor into the vapor generator so that only the non-condensable gases were removed, was also turned on. In the meantime, the auxiliary chiller (AD07R-40-A11B, Polyscience), integrated with the auxiliary brazed plate heat exchanger along the line of condensate return line, was turned on as well and set to 15 °C.

Once the coolant water flow in all chillers reaches steady condition, and after ensuring the vapor supply valve was fully closed, the vapor generator was filled with approximately 4 liters of the mixture of interest through the vapor discharge valve. The rope heater around the vapor generator was turned on with the heater power controller set to maximum output to boil the liquid rigorously. The temperature within the vapor generator was monitored by the installed RTD probe. During the boiling process, the vapor discharge valve was open to remove the dissolved non-condensable gases from the liquid mixtures. Once boiling was achieved and the temperature within the vapor generator was > 98 °C for at least 10 minutes, the vapor discharge valve was closed.

Prior to beginning the condensation experiments, the DSLR camera was turned on for visual recording of the sample during condensation. Afterwards, the power regulator of the rope vapor generator was manually adjusted to control the pressure inside the vapor generator and the vapor supply valve was slowly opened until the desired saturation pressure in the condensation chamber was reached. Quasi-steady state conditions were typically reached after 2 minutes of full operation. Upon completion of the experimental run, the heaters were first turned off while the cooling water was kept running for several hours to ensure proper cooldown of the system.

## S9. Experimental calculations and error propagation

### Condensation heat transfer coefficient

The heat transfer coefficient was obtained by first calculating the heat flux ($q^{"}$) through the condensation copper block using a one dimensional (1D) Fourier heat conduction approximation given by:

$$q^{"} = -k\frac{dT}{d\mathrm{x}} = \frac{Q}{A_1}, \tag{S6}$$

where $k$ is the thermal conductivity of copper (401 W/m.K) and $dT/d\mathrm{x}$ is the measured temperature gradient along the copper block. We thermally insulated the copper block using a custom-made PTFE-sleeve, which allows us to utilize the 1D heat transfer approximation. The three thermistors located at pre-determined and equally spaced locations in the copper block measured the temperature, from which the temperature gradient $dT/d\mathrm{x}$ was calculated, with $A_1$ denoting the corresponding cross-sectional area (see Figure S10(a)) and $Q$ the total heat transfer rate. The thermal resistance network represents the total resistance of the test rig, which included the copper sample, thermal paste and copper block. The thickness of the thermal paste was taken as 5 $\mu$m and the corresponding thermal conductivity was 4.7 W/m.K. To verify the validity of the 1D heat conduction assumption, the temperature gradients between adjacent thermistors were calculated from the experimental measurements and are plotted in Figure S10(b). The close agreement among these gradients indicates a nearly linear temperature profile across the sample, thereby confirming that the 1D heat conduction approximation is appropriate.

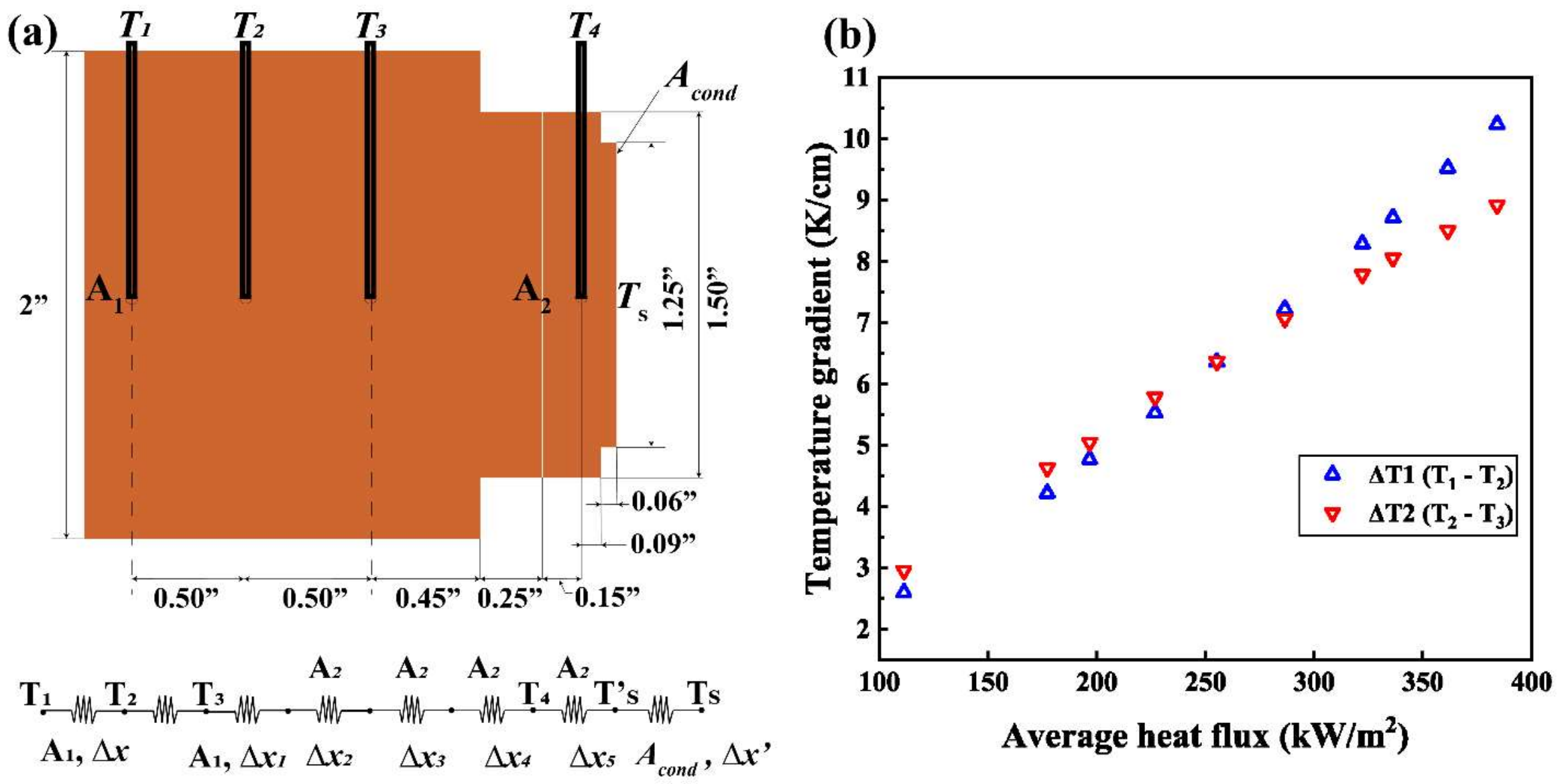

**Figure S10.** *(a) Side-view schematic of condensation copper block with thermistor hole placement and thermal resistance network. (b) A representative measurement showing the temperature gradient per unit distance between two consecutive thermistors.*

Based on the energy balance, the condensation heat flux ($q^{"}_{cond}$) on the surface area ($A_{cond}$) was calculated as:

$$q^{"}_{cond} = \frac{Q}{A_{cond}} = \frac{q^{"}A_1}{A_{cond}}. \qquad (S7)$$

The temperature of the condensing surface ($T_s$) could then be determined by assuming 1D heat conduction through the condensing block, as:

$$T_s = T'_s + \frac{q^{"}_{cond}\Delta x'}{k}, \qquad (S8)$$

where $T_4$ is the temperature measured by the thermistor close to the condensing surface and $\Delta x'$ is the distance between the condensing surface and the location of the thermistor.

The surface subcooling temperature ($\Delta T$) is defined as the temperature difference between the saturated vapor ($T_{sat}$) inside the chamber and the temperature of the condensing surface ($T_s$),

$$\Delta T = T_{sat} - T_s. \qquad (S9)$$

Here, $T_{sat} = T_{sat}(P_{sat})$ is taken as the saturation temperature of pure water at the measured chamber pressure $P_{sat}$.

The condensation heat transfer coefficient ($h_c$) is defined as the ratio of condensation heat flux to surface subcooling, which can be expressed as:

$$h_c = \frac{q^{"}_{cond}}{\Delta T}. \quad (S10)$$

**Error propagation analysis**

To quantify the uncertainty, $\delta(h_c)$, of our measured condensation heat transfer coefficients ($h_c$), we combined the known instrumental uncertainties associated with each measured quantity used for calculating the value of $h_c$. Because $h_c$ is determined from the measured heat flux ($q^{"}_{cond}$) and the measured subcooling ($\Delta T$), the overall systematic uncertainty of the condensation heat transfer coefficients follows from the propagating the individual errors in these terms, as expressed below [15]:

$$\delta(h_c) = \sqrt{\left(\frac{\partial h_c}{\partial q^{"}_{cond}}\delta(q^{"}_{cond})\right)^2 + \left(\frac{\partial h_c}{\partial \Delta T}\delta(\Delta T)\right)^2}. \quad (S11)$$

Based on the error propagation, the uncertainty of heat flux $\delta(q)$ was determined by:

$$\delta(q) = \sqrt{\left(\frac{\partial q^{"}_{cond}}{\partial T}\delta(T)\right)^2 + \left(\frac{\partial q^{"}_{cond}}{\partial x}\delta(x)\right)^2}. \quad (S12)$$

The uncertainty of the temperature measurements in the condensation block is subject to a $\pm 0.4$ ℃ error associated with the thermistor's specification. Additionally, the uncertainty associated with the distance between the pre-drilled holes for the thermistors in the condensation block, $\delta(x)$, was given by the machine shop to be $\pm$ 0.003".

The uncertainty of the surface subcooling temperature, $\delta(\Delta T)$, was determined by:

$$\delta(\Delta T) = \sqrt{\left(\frac{\partial \Delta T}{\partial T}\delta(T)\right)^2 + \left(\frac{\partial \Delta T}{\partial P}\delta(P)\right)^2}. \quad (S13)$$

Lastly, the uncertainty associated with the subcooling temperature difference (calculated according to eq. (S13) can be attributed to the thermistor error ($\pm 0.4$ ℃) combined with the uncertainty in the saturation temperature, $T_{sat}(P_{sat})$, which is directly tied to the uncertainty of the pressure transducer measurement ($\delta(P)$). Our high-pressure transducer (728A13TGA2FA, MKS) has a rated accuracy of $\pm 0.4\%$ given by the manufacturer, allowing us to determine the corresponding saturation temperature. The error bars reported in Figure 6 of the main manuscript were obtained by combining the subcooling temperature error ($\delta(\Delta T)$ with heat flux error ($\delta(q''_{cond})$) according to eq. (S11). Table S3 summarizes the uncertainties associated with each measured quantity.

*Table S3. Uncertainties corresponding to the direct experimental measurements*

| Experimental measurements | Uncertainty ($\pm$) |
|---|---|
| Calibrated thermistors | 0.4K |
| Saturated vapor pressure ($P_{Sat}$) | 0.5% |
| Length measurement ($x$) | 0.003” |

## S10. Filmwise condensation model and setup validation

To estimate the heat transfer coefficient of filmwise condensation on a smooth vertical flat Cu substrate, the classical Nusselt model was used, given by [16]:

$$h_{Nu} = 0.943 \left( \frac{\rho_l(\rho_l - \rho_v) g h'_{fg} k_l^3}{\mu_l D \Delta T} \right)^{\frac{1}{4}}, \tag{S14}$$

$$h'_{fg} = h_{fg} + c_{p,l} \Delta T, \tag{S15}$$

where $g$ is the gravitational acceleration ($9.81\ m/s^2$), $\rho_l$ is the liquid density, $\rho_v$ is the vapor density, $D$ is the diameter of the condensing substrate, $\mu_l$ is the dynamic viscosity of the condensate, $k_l$ is the thermal conductivity of the condensate, $h'_{fg}$ is the modified latent heat of vaporization accounting for the change in specific heat of condensate, and $c_{p,l}$ is the specific heat of the condensate.

An example of Marangoni and filmwise condensation heat transfer coefficients as a function of condenser subcooling temperature yielded from the above model is plotted in Figure 6 of the main manuscript under typical experimental conditions. The corresponding enhancement, defined as the ratio of the measured Marangoni condensation heat transfer to the filmwise prediction of Eq. (S15), is shown as a function of condenser subcooling in Figure S12.

Additionally, to validate the experimental setup, filmwise condensation of pure water was conducted and the measured heat transfer coefficients were compared against the theoretical predictions from the Nusselt model, which showed good agreement (see Figure S11).

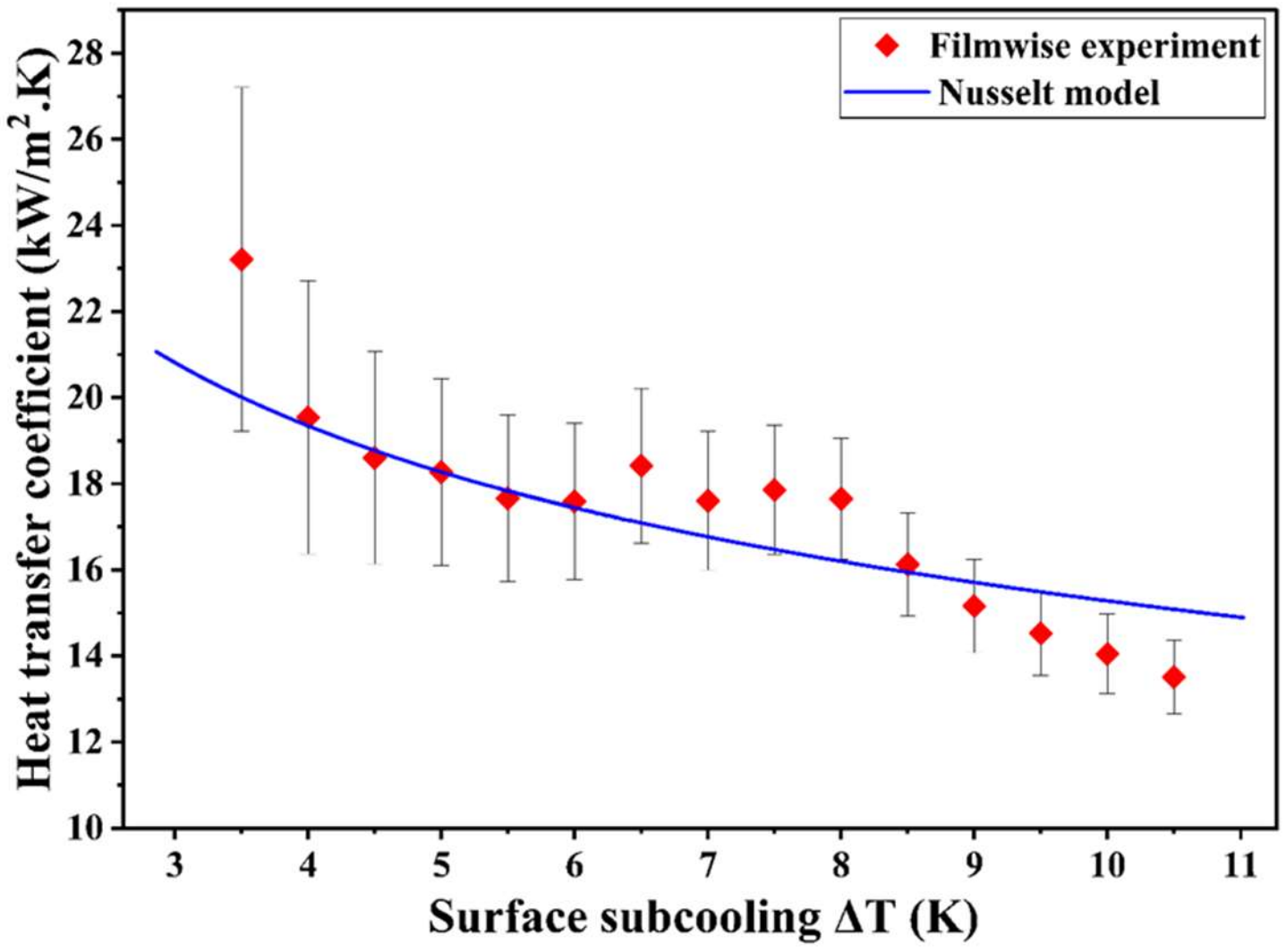

**Figure S11**. *Validation of experimental setup. Measured heat transfer coefficient as a function of surface subcooling compared with classical Nusselt theory for filmwise condensation of pure water.*

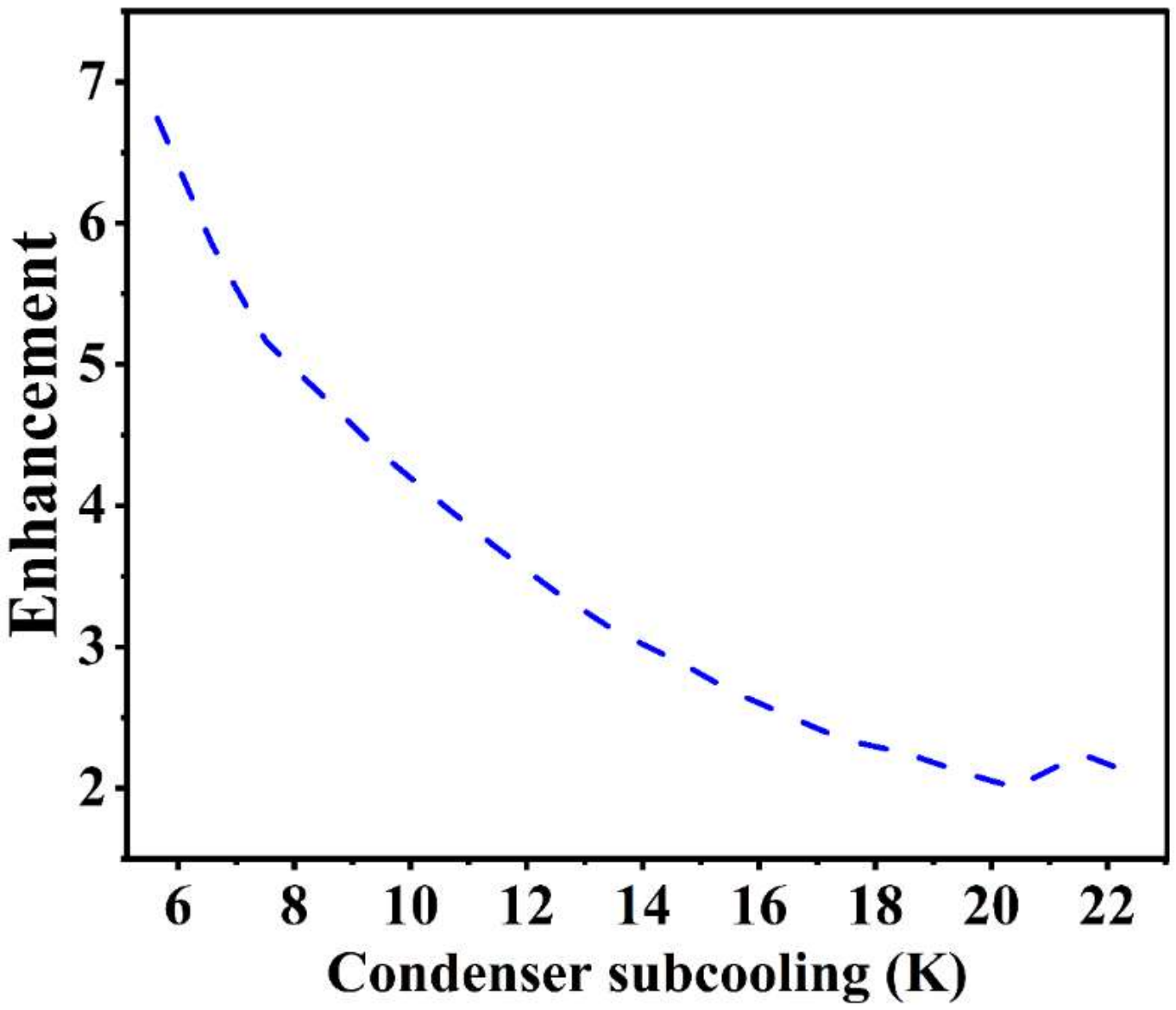

**Figure S12**. *Plot of Marangoni condensation enhancement as a function of condenser subcooling. The enhancement is the ratio of experimental Marangoni condensation heat transfer coefficient to filmwise condensation heat transfer coefficient given by Eq. S15.*

## S11. Time-resolved IR imaging of Marangoni dropwise condensation

Figure S13 shows time-lapse infrared (IR) images obtained during the Marangoni dropwise condensation of a water-EG mixture on the clean smooth copper surface during the initial transient condensate formation stage (see SM, video S3). As expected, when the vapor mixture contacts the condensing surface, small discrete droplets form and rapidly grow by coalescing with the neighboring droplets (see Figure S13(a-e)). Once the size of the droplets is large enough, the droplets are removed from the condensing surface by gravity (see Figure S13(f)). The same behavior is observed during the quasi-steady state operation (see Figure S14 and SM, video S4).

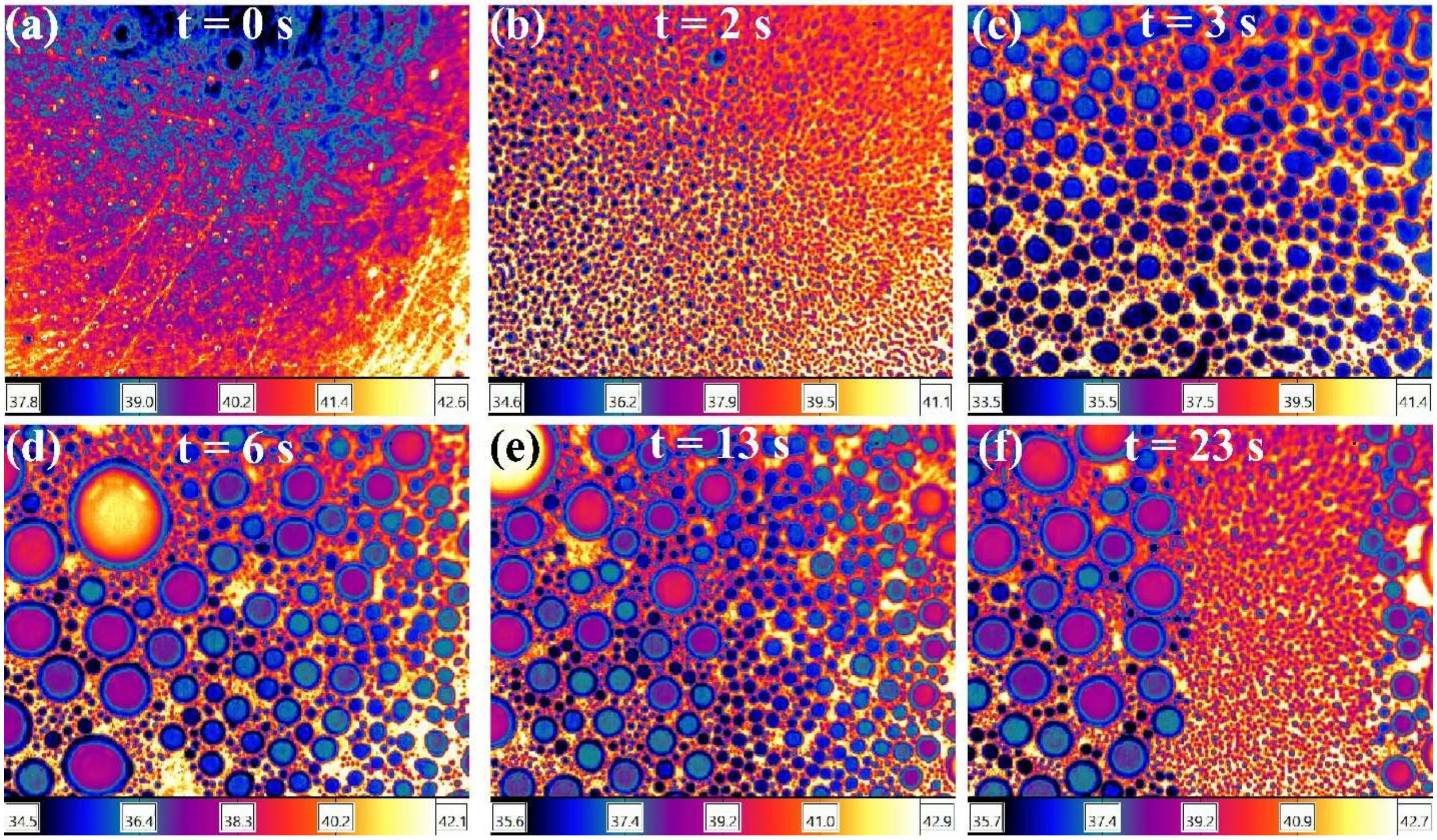

**Figure S13.** *Time-lapse IR images during the initial stages of water-EG mixture condensation.*

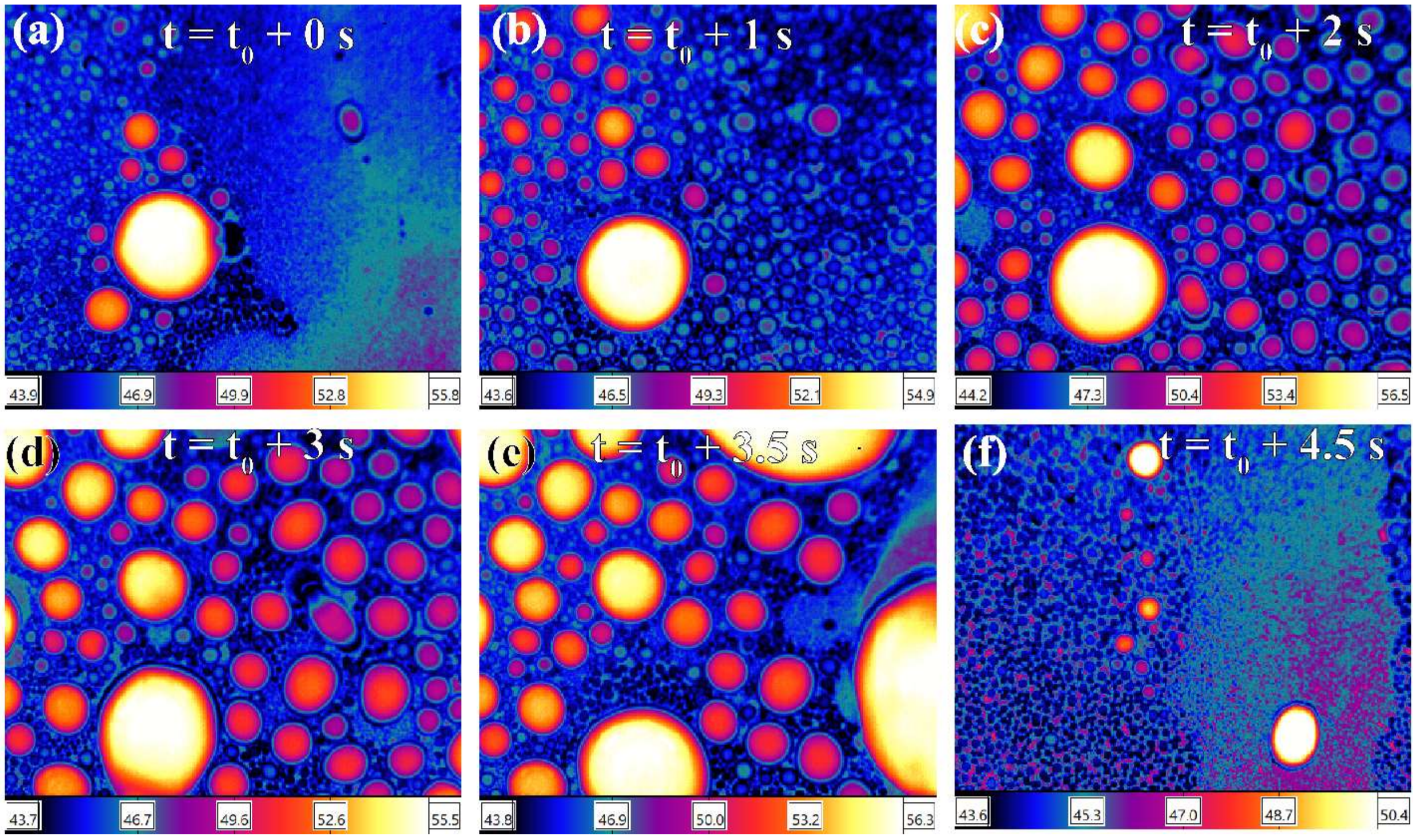

**Figure S14.** *Time-lapse IR images during the quasi-steady state condensation of a water-EG mixture.*